\documentclass{jfm}
\usepackage{graphicx}
\usepackage{epstopdf,epsfig,subfloat,subfig}

\usepackage{amsfonts}
\usepackage{amsmath}
\usepackage{amssymb,color,mathtools}
\usepackage{bm}% bold math

\usepackage{lineno}

\usepackage{hyperref}
\hypersetup{
	colorlinks=true,
	linkcolor=black,
	filecolor=black,      
	urlcolor=black,
	citecolor=black, 
}

\shorttitle{Extraction of the translational Eucken factor of molecular gas}
\shortauthor{Lei Wu, Haihu Liu and Wim Ubachs}

\title{Extraction of the translational Eucken factor from light scattering by molecular gas }
\author{
	Lei Wu\aff{1}
	\corresp{\email{wul@sustech.edu.cn}
	}, 
Haihu Liu\aff{2}
and Wim Ubachs\aff{3}
}

\affiliation{
	\aff{1}Department of Mechanics and Aerospace Engineering, Southern University of Science and Technology, Shenzhen 518055, China
	\aff{2}School of Energy and Power Engineering, Xi’an Jiaotong University, Xi’an 710049, China 
	\aff{3}Department of Physics and Astronomy, LaserLaB, Vrije Universiteit, De Boelelaan 1081, Amsterdam 1081 HV, The Netherlands 
}

\begin{document}

\maketitle

%\linenumbers

\begin{abstract}
	Although the thermal conductivity of molecular gases can be measured straightforwardly and accurately, it is difficult to experimentally determine its separate contributions from the translational and internal motions of gas molecules. Yet this information is critical in rarefied gas dynamics as the rarefaction effects corresponding to these motions are different. In this paper, we propose a novel methodology to extract the translational thermal conductivity (or equivalently, the translational Eucken factor) of molecular gases from the Rayleigh-Brillouin scattering (RBS) experimental data. From the numerical simulation of the~\cite{LeiJFM2015} model we find that, in the kinetic regime, in addition to bulk viscosity, the RBS spectrum is sensitive to the translational Eucken factor, even when the total thermal conductivity is fixed. Thus it is not only possible to extract the bulk viscosity, but also the translational Eucken factor of molecular gases from RBS light scattering spectra measurements. Such experiments bear the additional advantage that gas-surface interactions do not affect the measurements. For the first time, bulk viscosities (due to the rotational relaxation of gas molecules only) and translational Eucken factors of $\operatorname{N_2}$, $\operatorname{CO_2}$ and $\operatorname{SF_6}$ are simultaneously extracted from RBS experiments.
\end{abstract}

\section{Introduction}

The gas kinetic theory developed by~\cite{Maxwell1867} and~\cite{Boltzmann1872} has turned out to be extremely successful in describing the rarefied gas dynamics of dilute gas, and it has found a wide range of applications in  space vehicle reentry, micro-electromechanical systems, shale gas transport, and so on. Especially, when the intermolecular potentials are known, transport coefficients of monatomic gases such as the shear viscosity and thermal conductivity, obtained from the Chapman-Enskog expansion of the Boltzmann equation~\citep{CE}, agree well with experimental data. However, when the Boltzmann equation was extended to the~\cite{WangCS} equation for molecular gases, the difficulty arose to accurately determine the transport coefficients, e.g. the translational and internal Eucken factors ($f_{tr}$ and $f_{int}$, respectively) that appear in  the thermal conductivity $\kappa$ of a molecular gas 
\begin{equation}\label{thermal_theory}
\frac{\kappa{M}}{\mu_s}=\frac{3R}{2}f_{tr}+\left(c_v-\frac{3R}{2}\right)f_{int}\equiv{c_v}{f_u},
\end{equation}
where $\mu_s$ is the gas shear viscosity, $M$ is the molar mass, $R$ is the universal gas constant, $c_v$ is the molar heat capacity at constant volume, and $f_u$ is the total~\cite{Eucken1913} factor. 

%\begin{equation}\label{thermal_theory}
%\frac{\kappa{m}}{\mu_s{k_B}}=\frac{3}{2}f_{tr}+\frac{(d+h)}{2}f_{int}\equiv\frac{3+d+h}{2}{f_u},
%\end{equation}

It is also difficult to determine $f_{tr}$ and $f_{int}$ experimentally, despite that the thermal conductivity and total Eucken factor can be measured straightforwardly. However, in rarefied gas dynamics, $f_{tr}$ is a very important parameter since the rarefaction effects corresponding to the translational and internal motions of gas molecules are different. For example, in thermal creep flows where the gas automatically moves from a cold region to a hot region, even in the absence of a pressure gradient~\citep{Reynolds1879,Maxwell1879}, the thermal slip coefficient is proportional to $f_{tr}$ only, rather than $f_u$~\citep{porodnov1978thermal,Loyalka1979Polyatomic,loyalka1982thermal}. Although $f_{tr}$ can be measured in thermal creep flows~\citep{Mason1963JCP,gupta1970analysis}, the result cannot be accurate as it is hampered by the inaccurate gas-surface interaction~\citep{sharipov2011data,WuStruchtrupJFM2017}. 
% like the Maxwell's boundary condition that was proposed empirically

Recent advances in RBS experiments of molecular gases provide an excellent method to retrieve information on thermodynamic properties of gases~\citep{Pan2002,Pan2004,Vieitez2010,CRBS_JCP,guziyu,barker2013,GuRBS2015}, where the gas-surface interaction is absent, i.e. only a local volume inside the gas cell is probed by laser light. This extraction of gas information (especially the bulk viscosity) is achieved by comparing the experimental RBS line shapes with theoretical ones. According to~\cite{Sugawara1968PoF}, the RBS line shape can be obtained by solving the linearised Boltzmann equation for monatomic gas and \cite{WangCS} equation for molecular gas. However, due to the complexity of Boltzmann-type collision operators, simplified kinetic models like the \cite{Hanson1967PoF} model, on which the \cite{Tenti1974} line shape models are based, were proposed. Nowadays, the Tenti-S6 model is regarded as the most accurate approach, and its open source Matlab code is available based on the work of~\cite{Pan2004}. Most attention has been paid to the extraction of bulk viscosity, which is related only to the rotational relaxation time as the vibrational modes, even when they are activated fully or partially, remain ``frozen" at high frequencies used in light scattering experiments, see detailed discussions in Section III by~\cite{CRBS_JCP}.

The Tenti-S6 model requires four input parameters: (i) the uniformity parameter $y$ which is controlled by the shear viscosity, pressure, temperature, laser wavelength, and the angle of scattering; (ii) the internal degrees of freedom; (iii) the rotational relaxation time which determines the bulk viscosity in RBS experiments; and (iv) the effective thermal conductivity $\kappa_e$ involved in the light scattering. Since the vibrational relaxation is ``frozen" in RBS experiment, the internal degrees of freedom only correspond to the rotational degrees of freedom. For the same reason, in the determination of $\kappa_e$ the contribution from the vibrational degrees of freedom should be removed~\citep{Yang2017CPL,Yang2018Quantitative,WangJ2019CP}. That is
\begin{equation}\label{kappa_e}
\kappa_e=\frac{\mu_s}{M}\left(\frac{3R}{2}f_{tr}+\frac{d_rR}{2}f_{int}\right)\equiv\kappa-\frac{\mu_sd_vR}{2M}f_{int},
\end{equation}
%\begin{equation}\label{kappa_e}
%\kappa_e=\kappa-\frac{\mu_s{k_B}}{2m}hf_{int}.
%\end{equation}
where $d_r$ is the number of rotational degrees of freedom and $d_v=2c_v/R-3-d_r$ is the number of all other internal degrees of freedom that are able to hold heat energy. However, this subtraction imposes an uncertainty on the value of $\kappa_e$, because $f_{int}$ and $f_{tr}$ cannot be determined accurately from previous theoretical and experimental works~\citep{Mason1962,Mason1963JCP,gupta1970analysis}. Therefore, in order to extract accurate values of the bulk viscosity, the rotational relaxation time and translational Eucken factor should be varied simultaneously in gas kinetic models to find the minimum residual between the theoretical and experimental RBS line shapes. This cannot be achieved in the Tenti model, where $f_{tr}$ is related to the rotational collision time such that it is very unlikely to obtain the correct values of bulk viscosity and $f_{tr}$ at the same time.

In this paper we will use the kinetic model proposed by~\cite{LeiJFM2015}, which not only allows an independent variation of bulk viscosity and translational Eucken factor, but also incorporates the physical velocity-dependent collision frequency of gas molecules, such that this kinetic model is reduced to the Boltzmann equation in the limit of infinite rotational relaxation time. We will show that it is possible to extract the bulk viscosity and the translational Eucken factor simultaneously from RBS experiments.

The remainder of this paper is organized as follows. In Section~\ref{GKM} we introduce the~\cite{LeiJFM2015} model for molecular gases and the corresponding numerical method to compute the RBS spectrum. In Section~\ref{GKM_numerical} we compare the accuracy of the~\cite{LeiJFM2015} model and the \cite{Tenti1974} model, while in Section~\ref{compare_exp} the bulk viscosities and translational Eucken factors of $\operatorname{N_2}$, $\operatorname{CO_2}$ and $\operatorname{SF_6}$ are extracted from recent RBS experiments. Conclusions and perspectives are presented in Section~\ref{conclusion}.

\section{Gas kinetic models}\label{GKM}

\begin{figure}
	\centering
	{\includegraphics[width=0.6\textwidth]{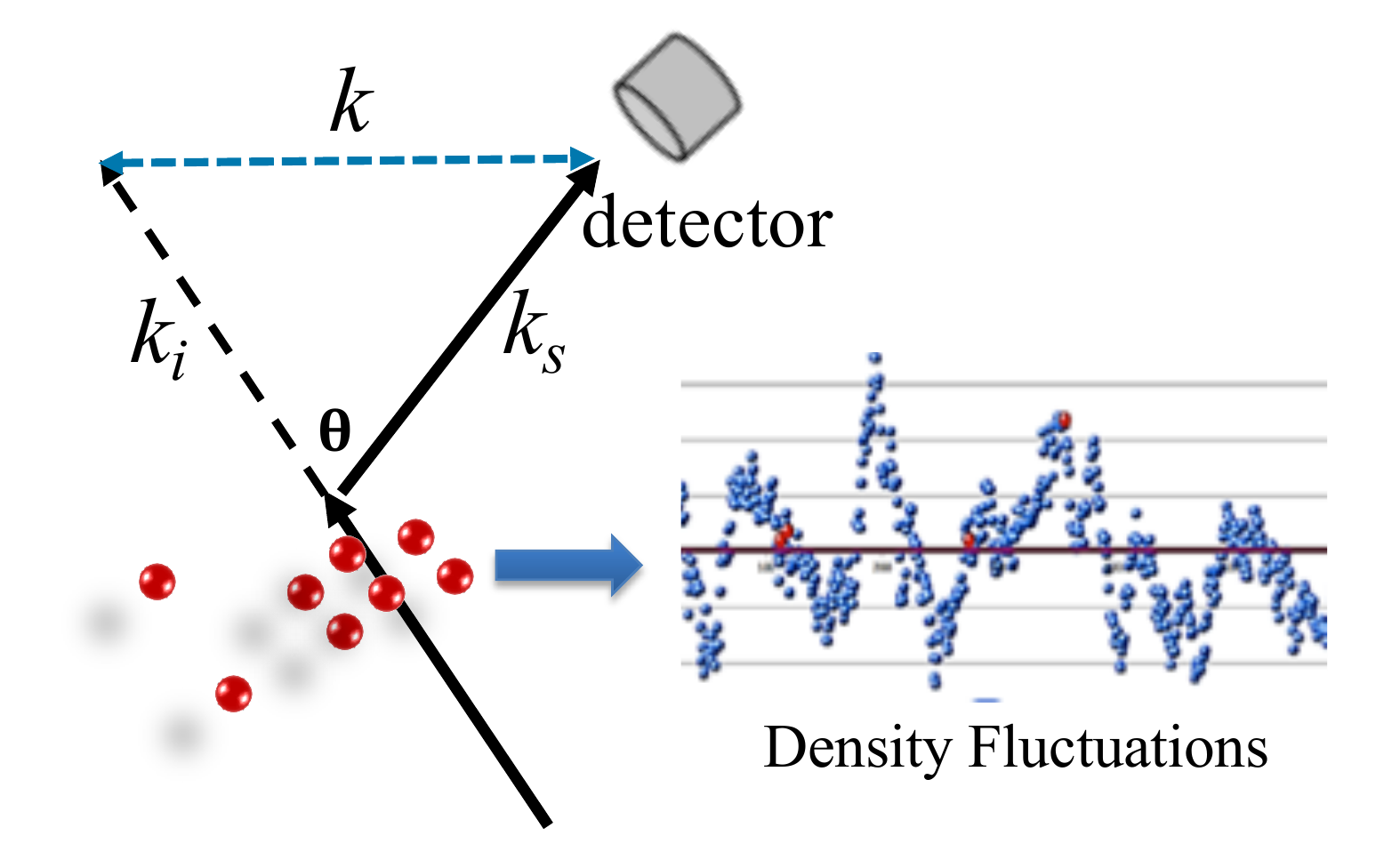}}
	\caption{Schematic of the spontaneous RBS process, where the light is scattered by the spontaneous density fluctuations in the gas. The vectors $k_i$ and $k_s$ represent the incident and scattered light wave momentum vectors, respectively, while $k$ represents the Brillouin scattering vector, in both directions for Stokes and anti-Stokes scattering.}
	\label{fig:RBS_demo}
\end{figure}

We consider the spontaneous RBS where the incident light with a wave vector $\bm{k}_i$ is scattered due to the thermal motion of gas molecules, see Figure~\ref{fig:RBS_demo}. Suppose the angle of scattering is $\theta$, the scattering wave propagates in the horizontal ($x_1$) direction with a wavenumber of
\begin{equation}
k_w=2|\bm{k}_i|\sin\left(\frac{\theta}{2}\right).
\end{equation} 

The spectrum of the scattered light is strongly influenced by the rarefaction parameter $\delta_{rp}$ (i.e. the ratio of the scattering wavelength $L=2\pi/k_w$ to the mean free path of gas molecules) used in the community of rarefied gas dynamics:
\begin{equation}\label{Kn_x}
\delta_{rp}=\frac{n_0k_BT_0}{\mu_s{(T_0)}v_m}L,
\end{equation} 
where $T_0$ and $n_0$ are the gas temperature and mean number density, respectively,  $v_m=\sqrt{{2k_BT_0}/{m}}$ is the most probable speed, $k_B$ is the Boltzmann constant, and $m$ is the mass of gas molecules. The rarefaction parameter is related to the uniformity parameter $y$ frequently used in RBS experiments and the~\cite{Tenti1974} model as 
\begin{equation}
\delta_{rp}=2\pi{y}.
\end{equation}

Although the Navier-Stokes equations are frequently used in fluid dynamics, they are only valid in the hydrodynamic regime at large values of $y$. To calculate the RBS line shape the gas kinetic theory should be adopted since the laser wavelength (scattering frequency) can be comparable to the mean free path (mean collision frequency) of gas molecules. In kinetic theory, the system state is described by the one-particle velocity distribution function in a seven-dimensional phase space, one dimension for time $t$, three dimensions for space $\bm{x}=(x_1,x_2,x_3)$, and three dimensions for molecular velocity space $\bm{v}=(v_1,v_2,v_3)$. All the macroscopic quantities such as density, velocity, temperature, pressure and heat flux are moments of this velocity distribution function. Governing equations for the velocity distribution function are introduced separately below for monatomic and molecular gases, i.e. for gases without and with internal degrees of freedom.

\subsection{Linearised kinetic equations for monatomic gases }

For monatomic gases, the evolution of velocity distribution function is governed by the Boltzmann equation. Since in spontaneous RBS density fluctuations are small, RBS spectra can be obtained by solving the following linearised Boltzmann equation~\citep{Sugawara1968PoF,lei_Jfm}:
\begin{equation}\label{LBE_s}
\frac{\partial {h}}{\partial{t}}+v_1\frac{\partial
	h}{\partial{x_1}}=\mathcal{L}(h)-\nu_{eq}(\bm{v}){h},
\end{equation}
where $h(t,x_1,\bm{v})$ is the velocity distribution function describing the deviation from the global equilibrium $f_{eq}(\bm{v})={\pi^{-3/2}}{\exp(-v^2)}$, $\mathcal{L}(h)$ is the gain part of the linearised Boltzmann collision operator $\mathcal{L}(h)=\int\int{}B [f_{eq}(\bm{v}')h({\bm{v}}')+h(\bm{v}')f_{eq}({\bm{v}}'_{\ast})-f_{eq}(\bm{v})h({\bm{v}}_\ast)]d\Omega d{\bm{v}}_\ast$,
%\begin{equation}\label{LBE_s_gain}
%\mathcal{L}(h)=\int\int{}B [f_{eq}(\bm{v}')h({\bm{v}}')+h(\bm{v}')f_{eq}({\bm{v}}'_{\ast})-f_{eq}(\bm{v})h({\bm{v}}_\ast)]d\Omega d{\bm{v}}_\ast,
%\end{equation}
and $\nu_{eq}(\bm{v})=\int\int B f_{eq}(\bm{v}_\ast)d\Omega{}d{\bm{v}}_\ast$ is the equilibrium collision frequency,
with $\bm{v}$ and $\bm{v}_\ast$ being the velocity of two molecules before the binary collision,  $\bm{v}'$ and $\bm{v}'_\ast$ the corresponding velocities after collision, and $d\Omega$ the solid angle of binary scattering. Generally speaking, the collision probability $B$, determined by the intermolecular potential, is a function of the relative collision speed $|\bm{v}-\bm{v}_\ast|$ and the deflection angle. For a Maxwellian gas where the intermolecular potential varies with the molecular separation $r$ as $r^{-5}$, $B$ is only a function of the deflection angle, and hence the equilibrium collision frequency is a constant. For a hard-sphere gas, $B$ is not a function of deflection angle but is proportional to the relative collision speed, such that $\nu_{eq}(\bm{v})$ increases with the molecular speed. If we consider the inverse power-law intermolecular potential, then the shear viscosity follows a power-law temperature scaling with the viscosity index $\omega$:
\begin{equation}\label{viscosity_index}
\mu_s(T)=\mu_s(T_0)\left(\frac{T}{T_0}\right)^\omega.
\end{equation} 
For hard-sphere and Maxwellian molecules we have $\omega=0.5$  and 1, respectively. The collisional behaviour of other gases usually falls between these two cases, and the viscosity index normally has value between 0.5 and 1~\citep{Bird1994}.

Note that in \eqref{LBE_s} the molecular velocity $\bm{v}$, spatial coordinate $x_1$, and time $t$ have been normalized by the most probable speed $v_m$, the scattering wavelength $L$, and $L/v_m$, respectively. Hence the collision probability $B$ is proportional to $\delta_{rp}$; details of this derivation and scaling have been presented by~\cite{Wu:2015yu}.

The initial perturbation in the spontaneous RBS process is a density impulse. Applying the Laplace-Fourier transform to~\eqref{LBE_s} in temporal-spatial directions, we obtain $2\pi{i}(f_s-v_1)\hat{h}=f_{eq}+\mathcal{L}(\hat{h})-\nu_{eq}\hat{h}$, where $i$ is the imaginary unit, $f_s$ is the frequency shift in the scattering process normalized by the characteristic frequency $v_m/L$, the hat denotes the Laplace-Fourier transform of the corresponding quantity, and $f_{eq}$ arises from the Laplace transform of initial density impulse. The spontaneous RBS spectrum $S$ is determined by:
\begin{equation}
S=\int \Re(\hat{h})d\bm{v},
\end{equation}
where $\Re$ is the real part of a variable and $\hat{h}$ is solved in the following iterative manner: 
\begin{equation}\label{iteration}
\hat{h}^{(j+1)}(\bm{v})=\frac{{f_{eq}(\bm{v})+\mathcal{L}(\hat{h}^{(j)})} } {2\pi{}i(f_s-v_1)+\nu_{eq}(\bm{v}) },
\end{equation}
with $j$ being the iteration step. The linearised Boltzmann collision operator can be solved by the fast spectral method~\citep{lei_Jfm}. Given the uniformity $y$ and frequency shift $f_s$, the iteration in \eqref{iteration} is terminated when the relative difference in the RBS spectrum $S$ between two consecutive iteration steps is less than $10^{-6}$. Starting from zero disturbance (i.e. $\hat{h}^{(0)}=0$), typically 20 iterations are sufficient to reach convergence at any value of $y$ when the general synthetic iterative scheme~\citep{SuArXiv2019} is used; and only a few seconds is needed to compute a line shape. Note that if the direct simulation Monte Carlo method~\citep{Bird1994} is used, several hours of computational time is needed to obtain one line shape, and the result often suffers from noise due to its stochastic nature.

The complexity of Boltzmann collision operator is usually much reduced by introducing gas kinetic models, where the essential physics such as the correct rates in the relaxation of shear stress and heat flux are kept. Here we introduce the~\cite{Shakhov1968} kinetic model, since it has been shown to produce results in good agreement with the direct simulation Monte Carlo method~\citep{Kun_Huang2011}; its linearised collision operator in this problem is 
\begin{equation}\label{shakhov_model}
\mathcal{L}_s=\delta_{rp}f_{eq}\left[\varrho+2u_1v_1+T\left(v^2-\frac{3}{2}\right)
+\frac{4}{15}q_1v_1\left(v^2-\frac{5}{2}\right)\right]-\delta_{rp}h,
\end{equation}
where the molecular number density deviating from the equilibrium value is
$\varrho=\int{h}d\bm{v}$, the flow velocity in the $x_1$ direction is $u_1=\int{v_1h}d\bm{v}$, the temperature deviating from the equilibrium value is $T=\frac{2}{3}\int{v^2}hd\bm{v}-\varrho$, and the heat flux in the $x_1$ direction is $q_1=\int{v_1v^2h}d\bm{v}-\frac{5}{2}u_1$. In this kinetic model, the collision frequency $\delta_{rp}$ is independent of the molecular speed, so it does not reflect the difference in different molecular interaction models (i.e. different value of viscosity index). Also, we find that when setting the rotational collision number to infinity~\citep{Pan2004}, the Tenti-S6 model is reduced to the linearised Skhahov model~\eqref{shakhov_model}.

\begin{figure}
	\centering
	\includegraphics[scale=0.33]{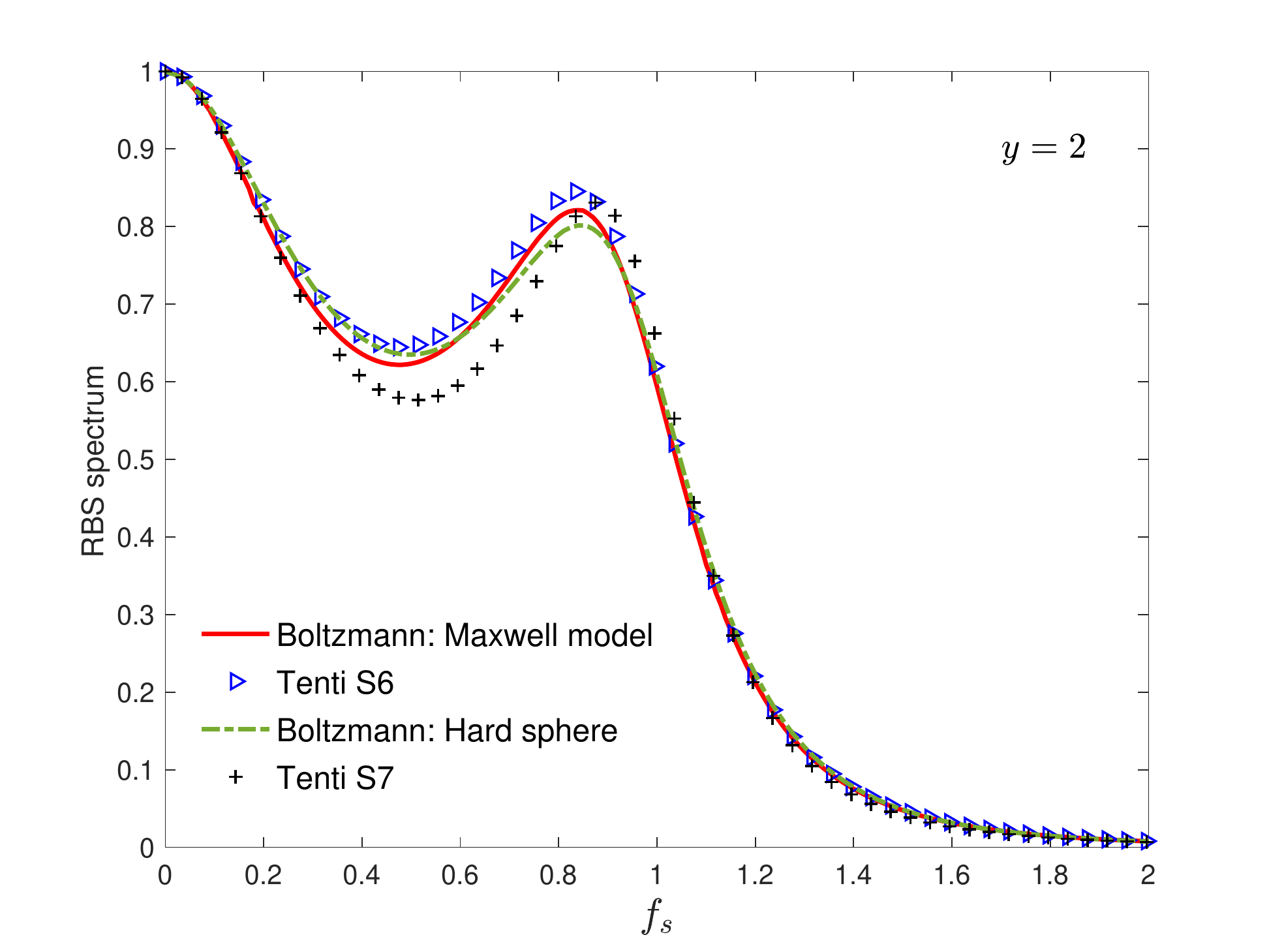}
	\includegraphics[scale=0.33]{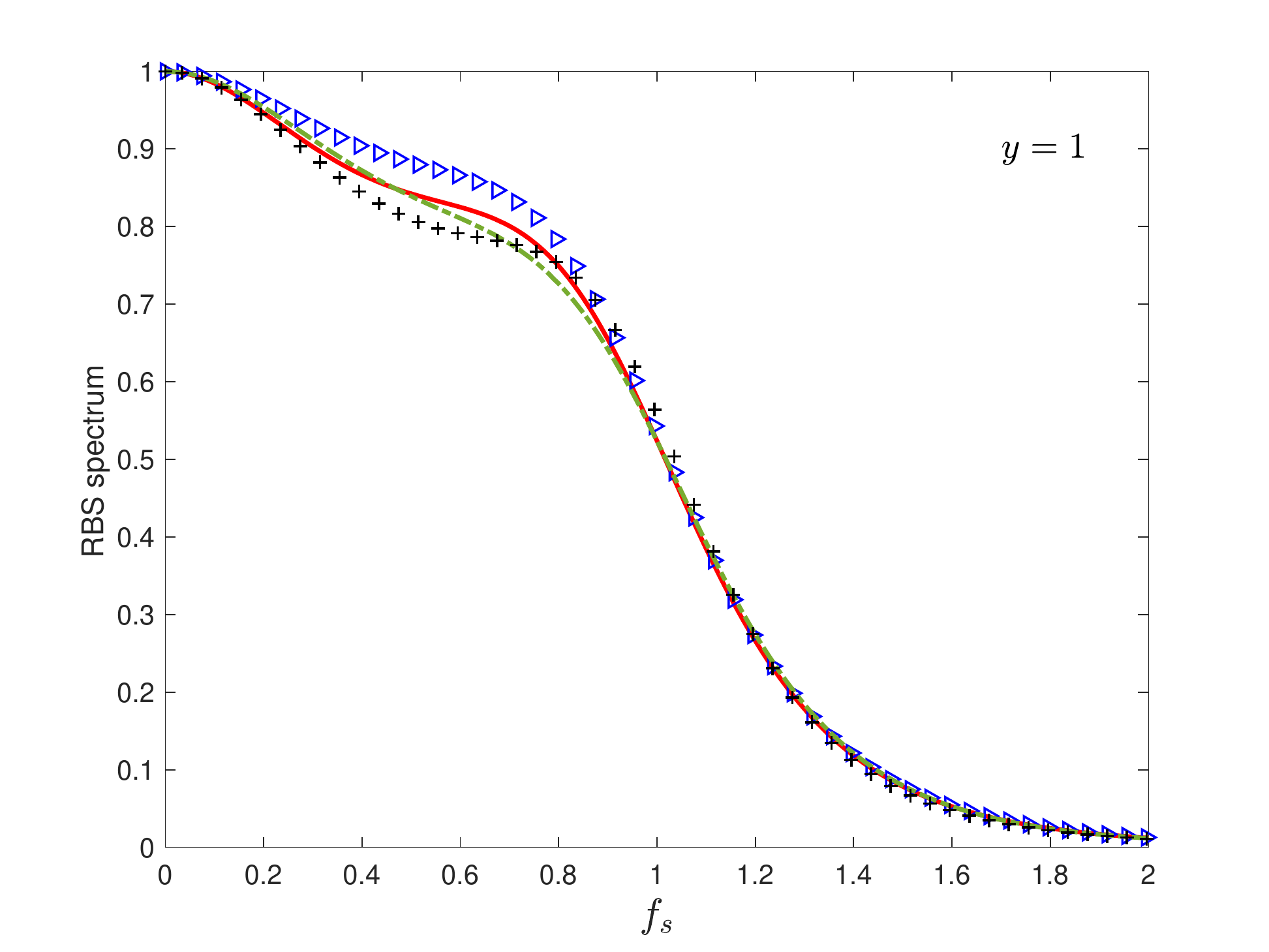}\\
	\includegraphics[scale=0.33]{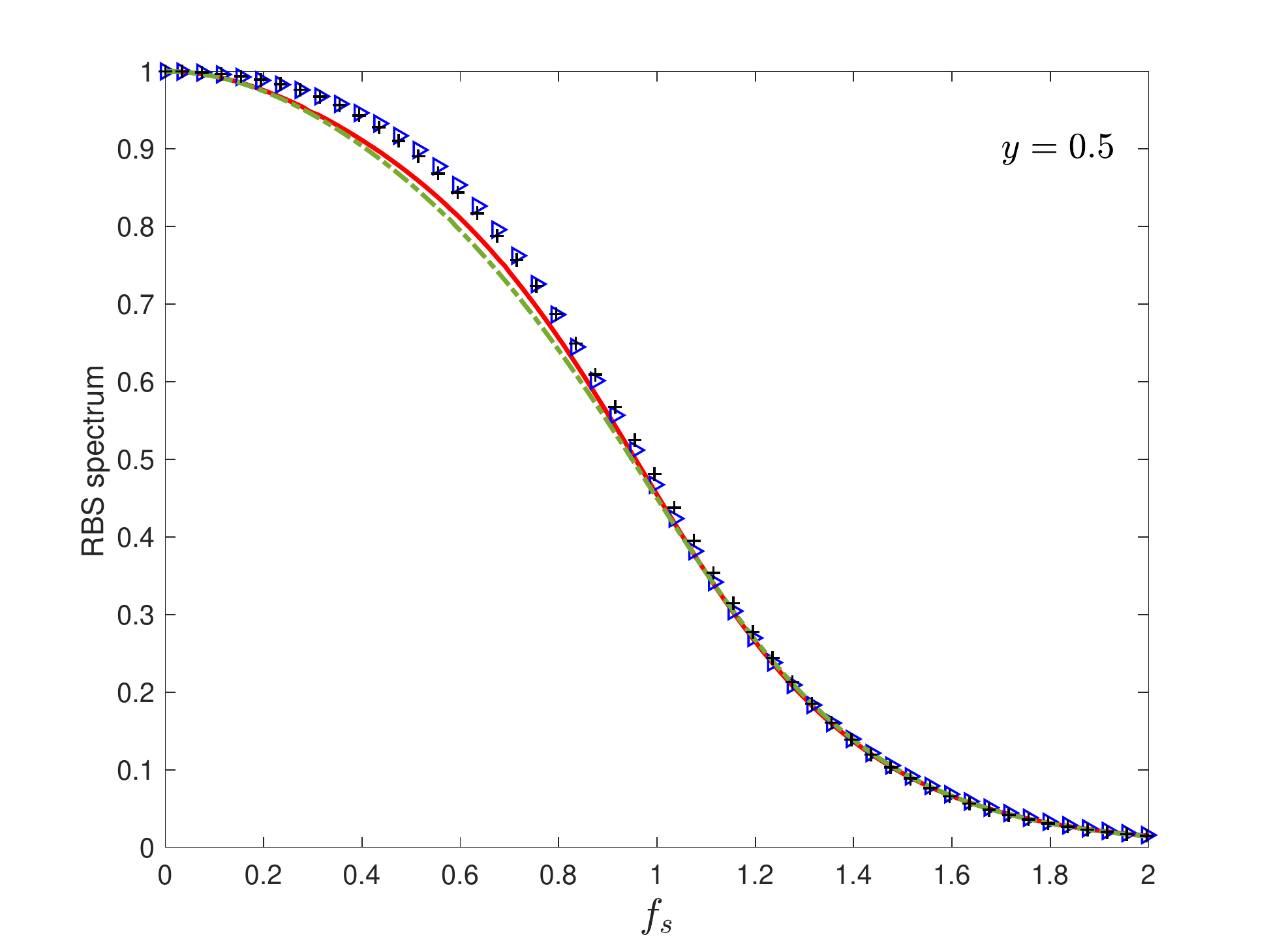}
	\includegraphics[scale=0.33]{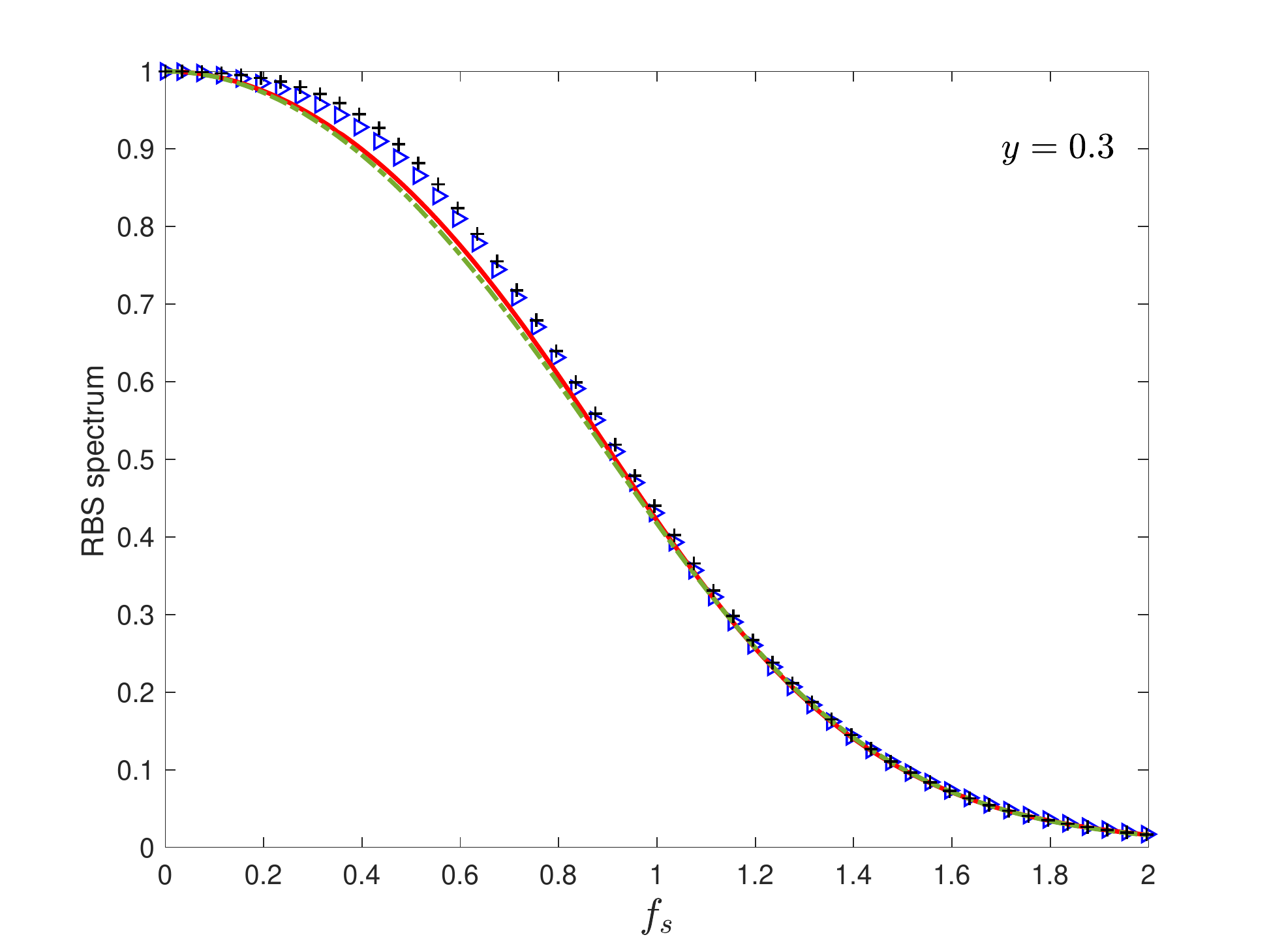}
	\caption{Computed RBS spectra from the linearised Boltzmann equation for Maxwellian and hard-sphere monatomic gases, as well as the Tenti-S6 and S7 line models~\citep{Pan2004}, where the frequency $f_s$ is normalized by $v_m/L$. In the Tenti models, in order to simulate the monatomic gas, we freeze the rotational relaxation by choosing a collision number of $10^{10}$. Each spectrum is normalized by its maximum value.  }
	\label{fig:monatomic_SRBS}
\end{figure}

% there must be something wrong fundamentally with the s7 model as its line shape is different from those of Boltzmann and S6 models even when $y$ is as large as 20. 

Figure~\ref{fig:monatomic_SRBS} compares the RBS spectra obtained from the linearised Boltzmann equation for Maxwellian and hard-sphere monatomic gases with calculations based on Tenti-S6 and Tenti-S7 models. When $y>5$, there is no difference between results from the present approach based on the Boltzmann equation and those from the Tenti-S6 model (not shown). This is because in the hydrodynamic regime both  Boltzmann equation and Tenti-S6 model yield the same Navier-Stokes equations under the Chapman-Enskog expansion. However, discrepancies in comparison with the Tenti-S7 model persist even at high pressures, or large $y$-values. For this reason the S7 model is discarded in the remainder of this paper. On the other hand, from the numerical simulations we find that the spectra from different line shape models (except the Tenti-S7 model) overlap visually when $y\lesssim0.2$. This is understandable, because the gas dynamics are effectively collisionless when $y$ approaches zero, and all kinetic models yield the same Gaussian line shape. In the kinetic regime, especially when $y\sim1$, differences in line shapes of Maxwellian and hard-sphere monatomic gases as well as those from Tenti-S6 model are visible. This is due to the fact that in this regime high-order moments (higher than these in the Navier-Stokes equations) play a role and their dynamics (e.g. relaxation rates) in different kinetic models are different.

\subsection{Kinetic models for molecular gases}

Kinetic equations describing the dynamics of molecular gases are much more complicated than those for monatomic gases. In addition to translational motion, molecules exhibit internal relaxation with exchanges of energy between the translational and internal modes. These relaxations lead to several new transport coefficients including the bulk viscosity and internal Eucken factor; they also affect the translational Eucken factor, make its value deviate from $5/2$ of monatomic gases~\citep{Mason1962}. \cite{WangCS} were the first to extend Boltzmann's work on monatomic gas to molecular gases, by assigning velocity distribution functions to each individual energy level. This brings tremendous analytical and computational difficulties; therefore it is  necessary to develop kinetic models to simplify the Boltzmann-type collision operator. In RBS, the \cite{Hanson1967PoF} model, on which the \cite{Tenti1974} model is based, serves this purpose. 

%{So the measured bulk viscosity is only determined by the rotational collision time as
%\begin{equation}\label{bulk_theory}
%\mu_b=2p\tau_r\frac{d}{(3+d)^2}.
%\end{equation}
%Note that the rotational collision number in~\eqref{Eucken} is related to $\tau_r$ as $Z_{r,coll}=4p\tau_r/\pi\mu_s$.
%}

Details of the \cite{Hanson1967PoF} model in RBS application are given by~\cite{Tenti1974} and~\cite{Pan2004}. We only summarize the major results here. When the vibrational relaxation is ``frozen'', the bulk viscosity $\mu_b$ from the Tenti model is 
\begin{eqnarray}\label{bulk_viscosity}
\frac{\mu_b}{\mu_s}=\frac{2d_rZ}{3(d_r+3)}, %\equiv\frac{\pi{}dZ_{r,coll}}{2(d+3)^2},
\end{eqnarray}
where $Z$ is the rotational collision number such that in spatial-homogeneous state the rotational temperature $T_r$ relaxes to the total temperature $T$ as ${\partial{}T_r}/{\partial t}={(T-T_r)}/{Z\tau}$,
with $\tau=\mu_s(T_0)/n_0k_BT_0$ being the mean collision time of gas molecules (in this equation temperature and time have not been normalized). Note that $Z$ is exactly the same as $R_{int}$ defined by~\cite{Pan2004}. On the other hand, when the effective  thermal conductivity $\kappa_e$ defined in~\eqref{kappa_e} and the bulk viscosity  (controlled by $Z$) are known, the translational Eucken factor is uniquely determined as~\citep{Loyalka1979Polyatomic}
\begin{equation}\label{ftr_Tenti}
f_{tr}=\frac{5f_u^e+10Z}{5+4Z}, \quad \operatorname{with~} f_u^e=\frac{2m\kappa_e}{(3+d_r)\mu_s{k_B}},
\end{equation}
where $f_u^e$ is the effective  Eucken number.

However, since the interaction between molecules is not in all cases governed by a symmetric Maxwellian potential as assumed by~\cite{WangCS}, such a unique relation might not hold for all molecules, in particular for polar molecules. Therefore, a kinetic model for a molecular gas which allows the independent change of bulk viscosity and translational Eucken factor is needed. The model developed by \cite{LeiJFM2015} satisfies this requirement; it introduces two velocity distribution functions $h_0$ and $h_2$ related to the translational and rotational motions of gas molecules to describe the system state:
\begin{equation}\label{SRBS}
\begin{aligned}[b]
\frac{\partial{h_0}}{\partial {t}}+v_1\frac{\partial h_0}{\partial x_1}=\mathcal{C}_0\equiv&\mathcal{L}(h_0)-\nu_{eq}(\bm{v}){h_0} \\
+& \delta_{rp}\frac{f_{eq}}{Z}\left[(T-T_t)\left(v^2-\frac{3}{2}\right)+\frac{4(\omega_0-1)}{15}q_tv_1\left(v^2-\frac{5}{2}\right)\right], \\
\frac{\partial{h_2}}{\partial {t}}+v_1\frac{\partial h_2}{\partial x_1}=\mathcal{C}_2\equiv&\delta_{rp}\left[\frac{d_r}{2}T_rf_{eq}-h_2\right]
+ \frac{d_r\delta_{rp}}{2Z}(T-T_r)f_{eq}\\
+&\delta_{rp}\frac{{2(Z+\omega_1-1)(1-\delta)}}{Z} q_rv_1f_{eq}, 
\end{aligned}
\end{equation}
where the molecular density is $\varrho=\int{h_0}d\bm{v}$, the flow velocity is $u_1=\int{v_1h_0}d\bm{v}$,  the translational temperature is $T_t=\int{(2v^2/3-1)h_0}d\bm{v}$, the rotational temperature is $T_r=(2/d_r)\int{h_2}d\bm{v}$, the total temperature is $T=(3T_t+d_rT_r)/(3+d_r)$, the translational heat flux is $q_t=\int{(v^2-5/2)v_1h_0}d\bm{v}$, and the rotational heat flux is $q_r=\int{v_1h_2}d\bm{v}$. The two parameters $\omega_0$ and $\omega_1$ can be determined by setting the following translational and internal Eucken factors, obtained by applying the Chapman-Enskog expansion to~\eqref{SRBS},
\begin{equation}\label{thermal_conductivity}
\begin{aligned}[b]
f_{tr}=&\frac{5}{2}\left(1+\frac{1-\omega_0}{2Z}\right)^{-1}, \\ 
f_{int}=&\delta^{-1}\left[1+\frac{(1-\delta)(1-\omega_1)}{\delta{}Z}\right]^{-1},
\end{aligned}
\end{equation}
equal to the real values of molecular gases, where $\delta$ is the Schmidt number.

It should be noted that in the limit of infinite rotational collision number, i.e. $Z\rightarrow\infty$, the first equation in \eqref{SRBS} is reduced to the linearised Boltzmann equation \eqref{LBE_s} and $f_{tr}$ takes the value of $5/2$ for monatomic gases.

While the bulk viscosity is determined by $Z$ through~\eqref{bulk_viscosity}, the translational and internal Eucken factors can be controlled independently by adjusting the value of $\omega_0$ and $\omega_1$, respectively. This is one of the advantages of the Wu model~\eqref{SRBS}, while in the Tenti model $f_{tr}$ is uniquely determined by $Z$ through \eqref{ftr_Tenti}. On the other hand, when the linearised Boltzmann collision operator $\mathcal{L}(h_0)-\nu_{eq}(\bm{v}){h_0}$ in~\eqref{SRBS} is replaced by that of the Shakhov model~\eqref{shakhov_model}, i.e
\begin{equation}
\mathcal{L}_s=\delta_{rp}f_{eq}\left[\varrho+2u_1v_1+T_t\left(v^2-\frac{3}{2}\right)
+\frac{4}{15}q_{t}v_1\left(v^2-\frac{5}{2}\right)
\right]-\delta_{rp}h_0,
\end{equation}
the resultant kinetic model becomes the extended Rykov model~\citep{Rykov,LeiJFM2015}. Through the numerical simulation we find that, at the same values of $f_{tr}$ and $f_u^e$, the Tenti-S6 model and the extended Rykov model produce nearly the same results. Since for monatomic gas the linearised Boltzmann equation is more accurate than the Shakhov model, it may be concluded that for molecular gases the Wu model~\eqref{SRBS} is more reliable than the Tenti-S6 model. This is the second advantage of the Wu model.

Similar to the case of monatomic gases, the RBS spectrum $S=\int \Re(\hat{h}_0)d\bm{v}$ of a molecular gas can be obtained by solving the following equations iteratively:
\begin{equation}\label{iteration_srbs}
\begin{aligned}[b]
\hat{h}_0^{(j+1)}(\bm{v})=&\frac{f_{eq}+\hat{\mathcal{C}}_0^{(j)}+\nu{\hat{h}_0^{(j)}} }{2\pi{}i(f_s-v_1)+\bar{\nu}},\\
\hat{h}_2^{(j+1)}(\bm{v})=&\frac{\hat{\mathcal{C}}_2^{(j)}+\nu{\hat{h}_2^{(j)}} }{2\pi{}i(f_s-v_1)+\bar{\nu}},
\end{aligned}
\end{equation}
where the constant $\bar{\nu}=1.5\delta_{rp}(1+1/Z)$ is chosen to ensure the stability of iterations. A similar general synthetic iterative scheme~\citep{SuArXiv2019} can be developed to obtain a converged solution within about 20 iterations at any values of $f_s$ and $y$. This accurate and efficient numerical scheme enables us to extract the gas information from experimental data.

\section{Numerical comparison in molecular gas}\label{GKM_numerical}

In this section detailed comparisons between the Wu model and the Tenti-S6 model will be presented, where we will focus on the influence of the rotational collision number $Z$, the translational Eucken factor $f_{tr}$, and the ``effective'' thermal conductivity (or equivalently the effective Eucken factor $f_u^e$) on the RBS line shape. %The ``effective'' means that the vibrational thermal conductivity is excluded in the calculation of Eucken factors. 

\subsection{Role of rotational collision number}\label{rotational_role}

When the uniformity parameter $y$ is small, the gas dynamics is described by the collisionless Boltzmann equation so any internal relaxations are frozen. That is, the RBS spectrum of molecular gas is exactly the same as that of the monatomic gas, adopting a Gaussian shape. To study the influence of the rotational collision number on the RBS line shapes we need to choose larger values of $y$. We consider only the Wu model~\eqref{SRBS} here; since it allows the independent change of bulk viscosity and translational Eucken factor we can vary the value of $Z$ but keep $f_{tr}$ fixed. To this end we choose $f_{tr}=f_{int}=2$, as the influence of $Z$ on the RBS line shape will be similar for other values of $f_{tr}$ and $f_{int}$.

\begin{figure}
	\centering
	{\includegraphics[scale=0.33]{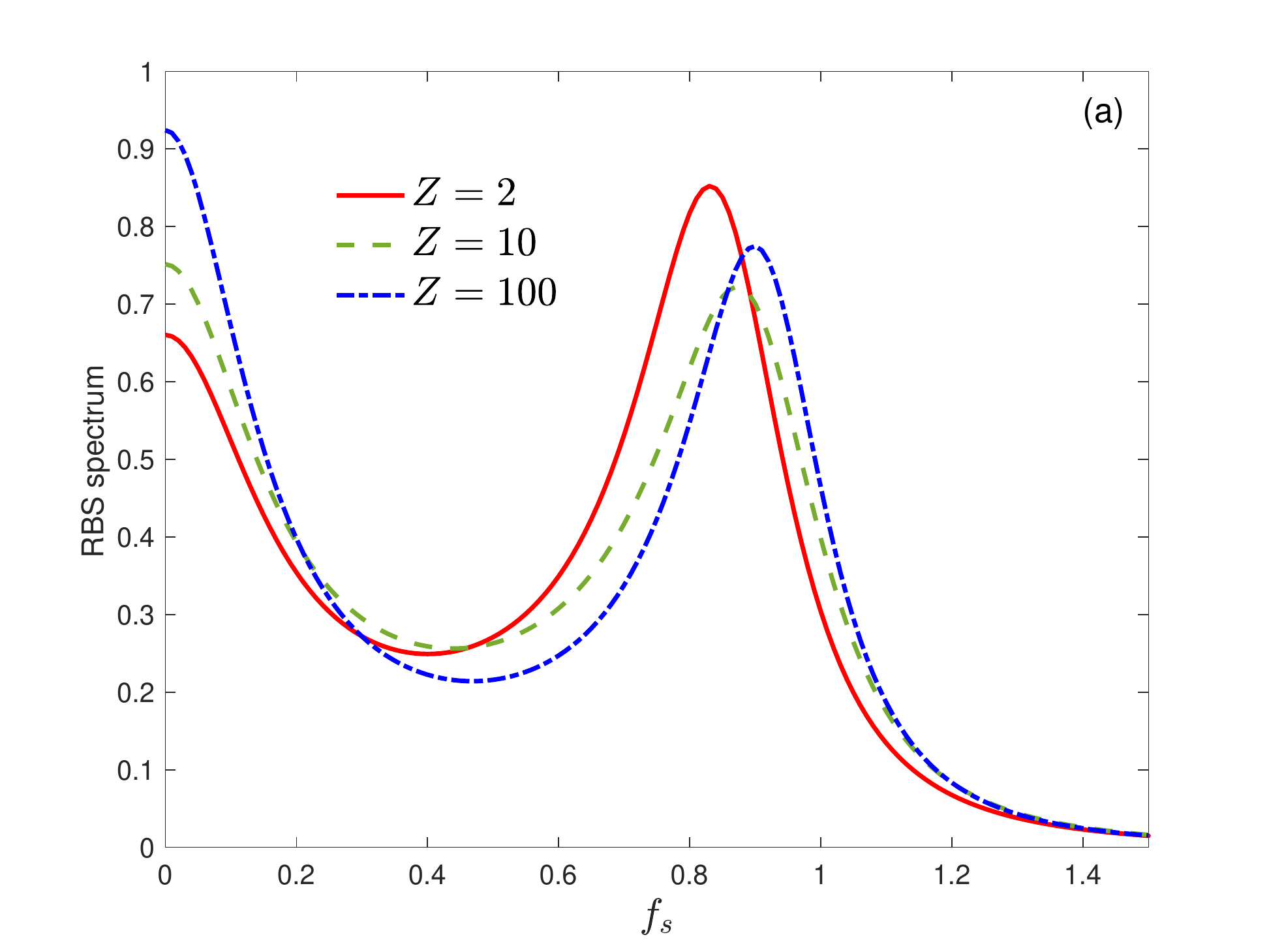}}
	{\includegraphics[scale=0.33]{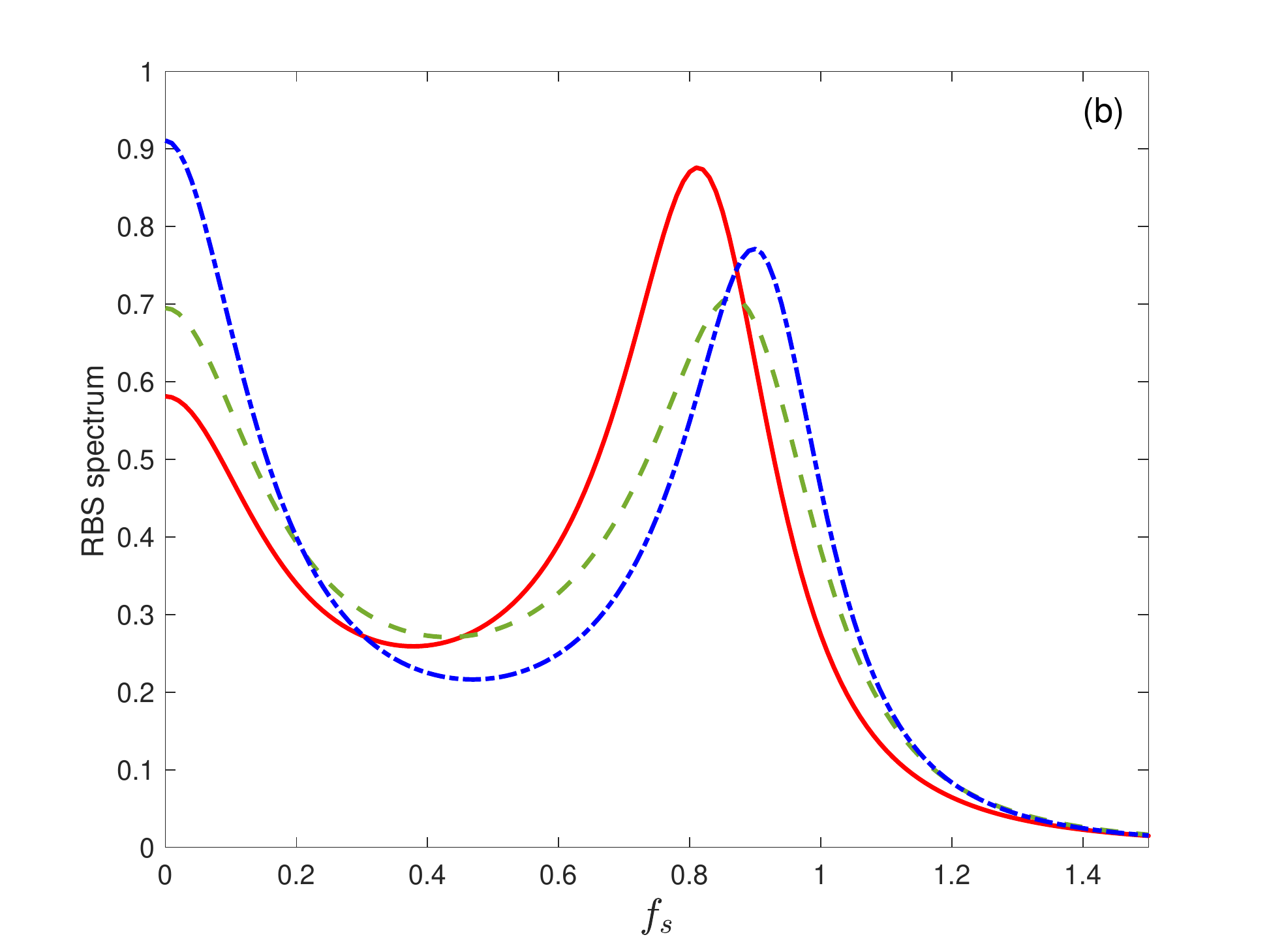}}
	\caption{Computed RBS spectra from the Wu model (2.12) showing the influence of rotational collision number $Z$ in the case of (a) a diatomic gas with $d_r=2$  and (b) a gas of nonlinear polyatomic molecules  with $d_r=3$, when the translational and rotational Eucken factors are $f_{tr}=f_{int}=2$, and the uniformity parameter is $y=4$. In this and following figures, if not further specified, the RBS spectrum is normalized by the area $\int_{-\infty}^{\infty}S(f_s)df_s$ of each line shape. }
	\label{fig:polyd2h1_rot}
\end{figure}

Figure~\ref{fig:polyd2h1_rot} shows the RBS spectra for $y=4$, where the influence of intermolecular potential is negligible, as discussed above near Figure~\ref{fig:monatomic_SRBS}. The spectrum is normalized by the area $\int_{-\infty}^{\infty}S(f_s)df_s$. It can be seen that the spectrum near the central Rayleigh peak increases with $Z$. Also, for $d_r=2$ (e.g. diatomic gases), the position of the side Brillouin peak shifts from $f_s=0.83$ to 0.90 when $Z$ increases from 2 to 100, while in the case of $d_r=3$ (i.e. nonlinear polyatomic gases) this peak moves from 0.81 to 0.90. Since this position is approximately determined by the sound speed normalized by the most probable speed $v_m$, it can be concluded that the rotational relaxation is gradually frozen such that the sound speed of the molecular gas increases from $\sqrt{(5+d_r)/2(3+d_r)}$ to that of a monatomic gas $\sqrt{5/6}$. Note that the collision number corresponding to vibrational modes is on the order of $10^4$ for CO$_2$ and SF$_6$, and even larger for other molecules~\citep{CRBS_JCP}. Therefore, this example implies that, when applying the Tenti-S6 model, the internal degrees of freedom should be the rotational degrees of freedom $d_r$, and the total thermal conductivity should only take into account the contribution from the translational and rotational motions. Indeed, in the analysis of experimental RBS spectra it was concluded that the vibrational relaxation was found to be frozen in various molecular gases~\citep{Yang2017CPL,Yang2018Quantitative,WangJ2019CP}.

\subsection{Role of translational thermal conductivity}\label{translational_role}

In~\eqref{thermal_theory} we see that the total thermal conductivity consists of the translational and internal parts, which are closely related to the translational motion and diffusion of gas molecules, respectively. In the hydrodynamic regime, the Navier-Stokes equations are valid and the line shape is determined by the total thermal conductivity. On the contrary, in the free-molecular regime, the RBS spectrum does not reflect a specific thermal conductivity, i.e. a gas with any value of thermal conductivity shares the same Gaussian line shape centred at $f_s=0$. The role of thermal conductivity in the kinetic regime, as an intermediate between these two limiting cases, is not a priori determined. Especially the case how the RBS line shape behaves under conditions of varying $f_{tr}$ but fixed $f_u^e$ and $\mu_b$ has never been studied.

\begin{figure}
	\centering
	\includegraphics[scale=0.33]{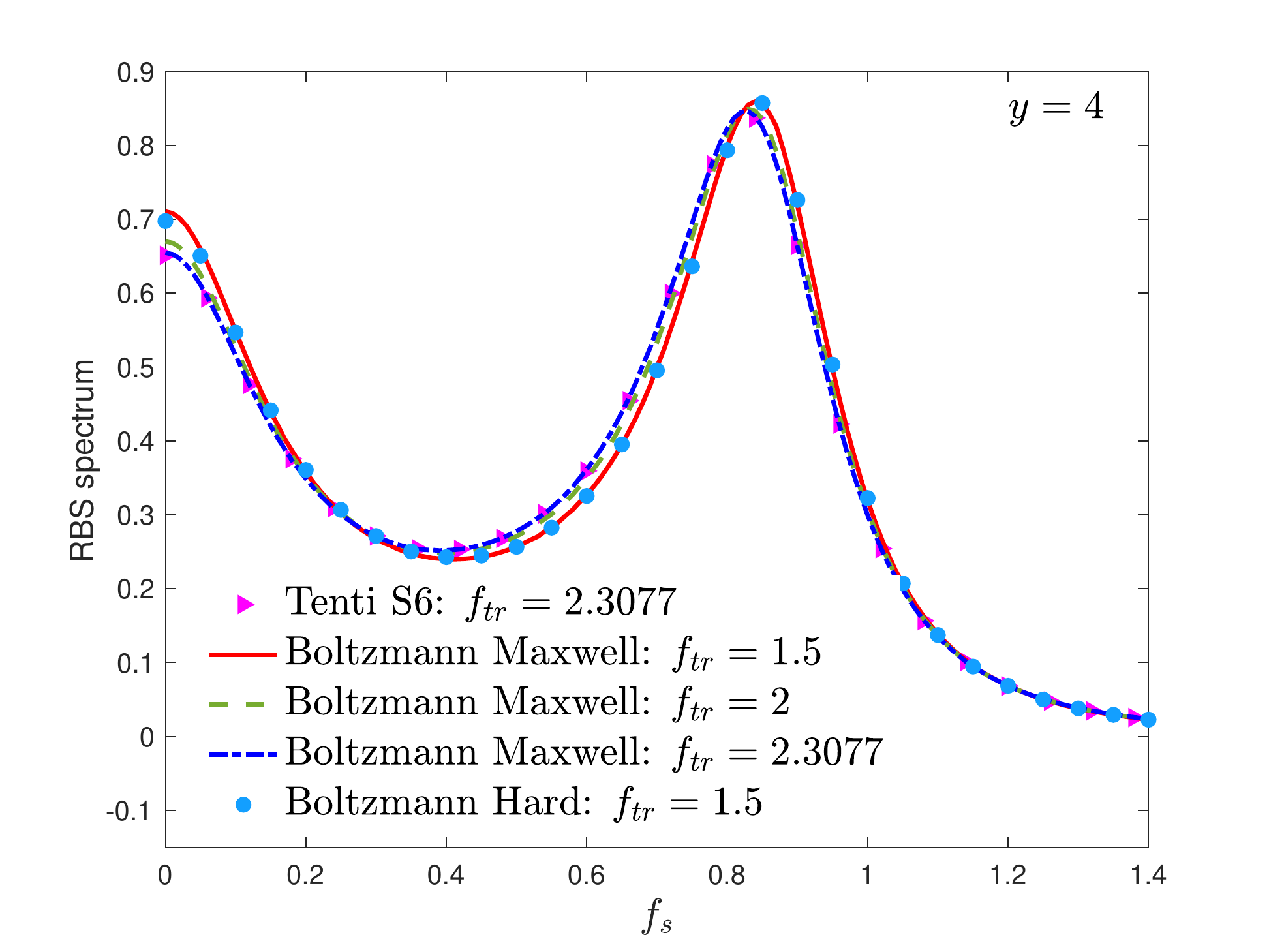}
	\includegraphics[scale=0.33]{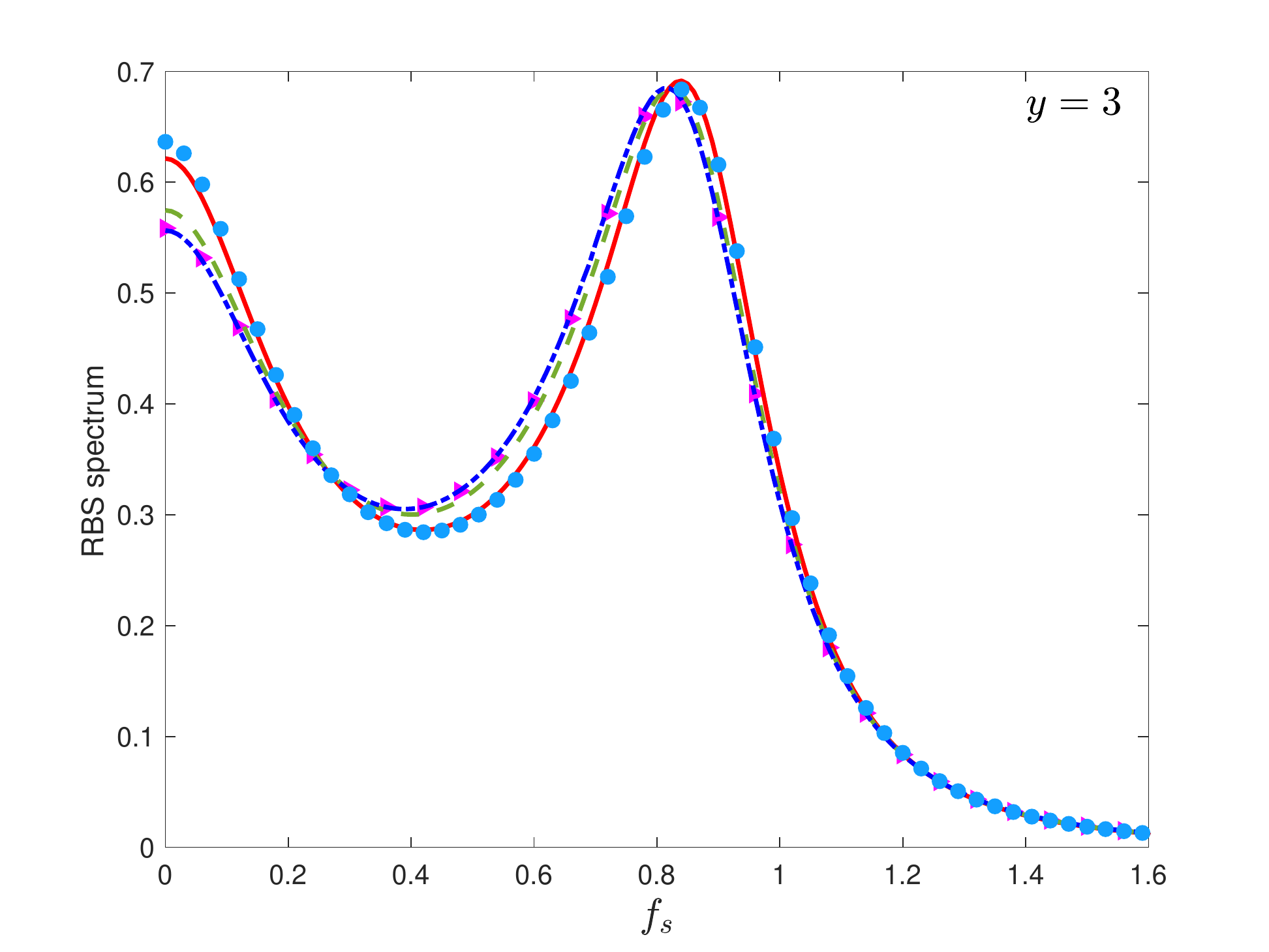}\\
	\includegraphics[scale=0.33]{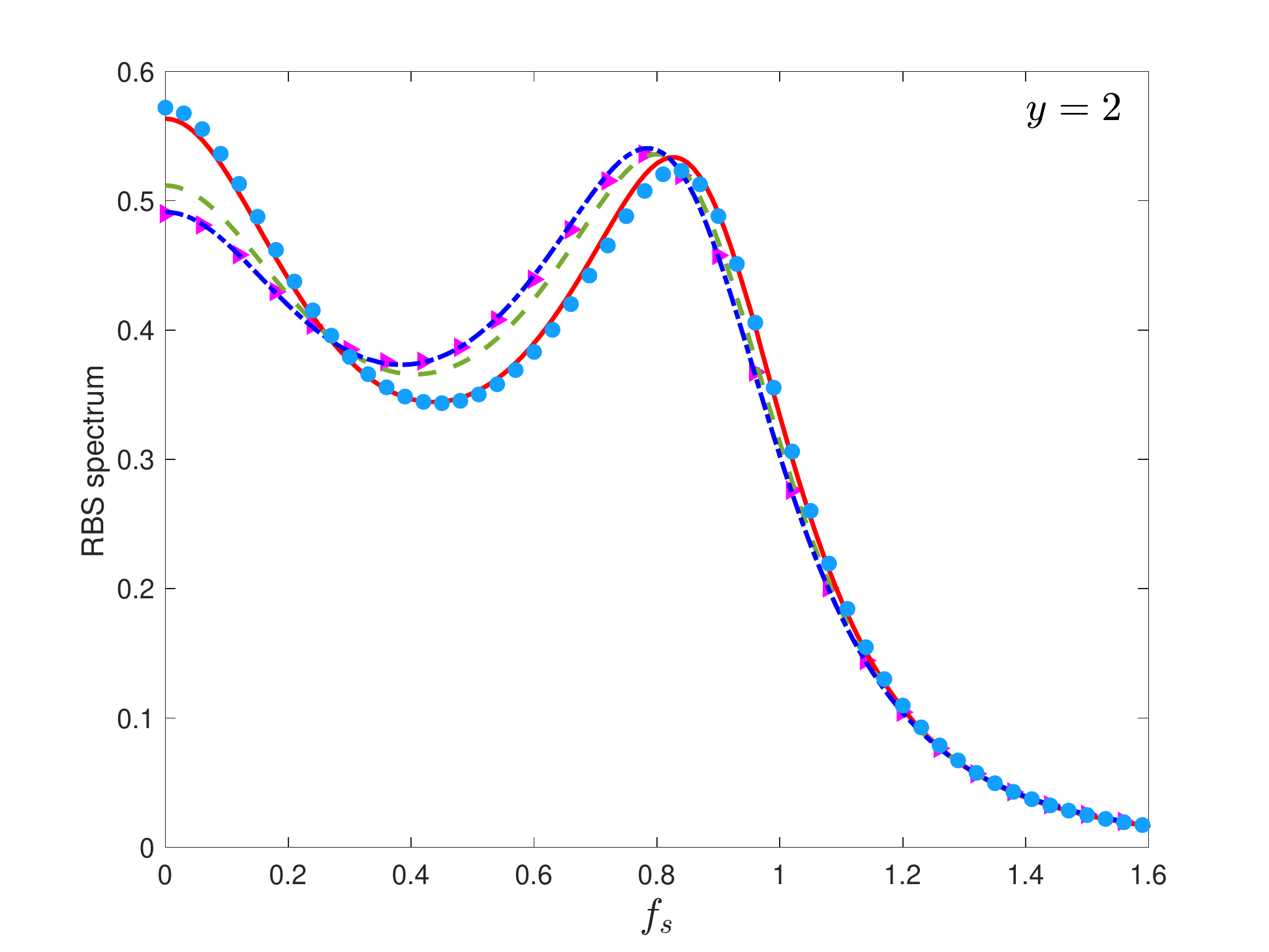}
	\includegraphics[scale=0.33]{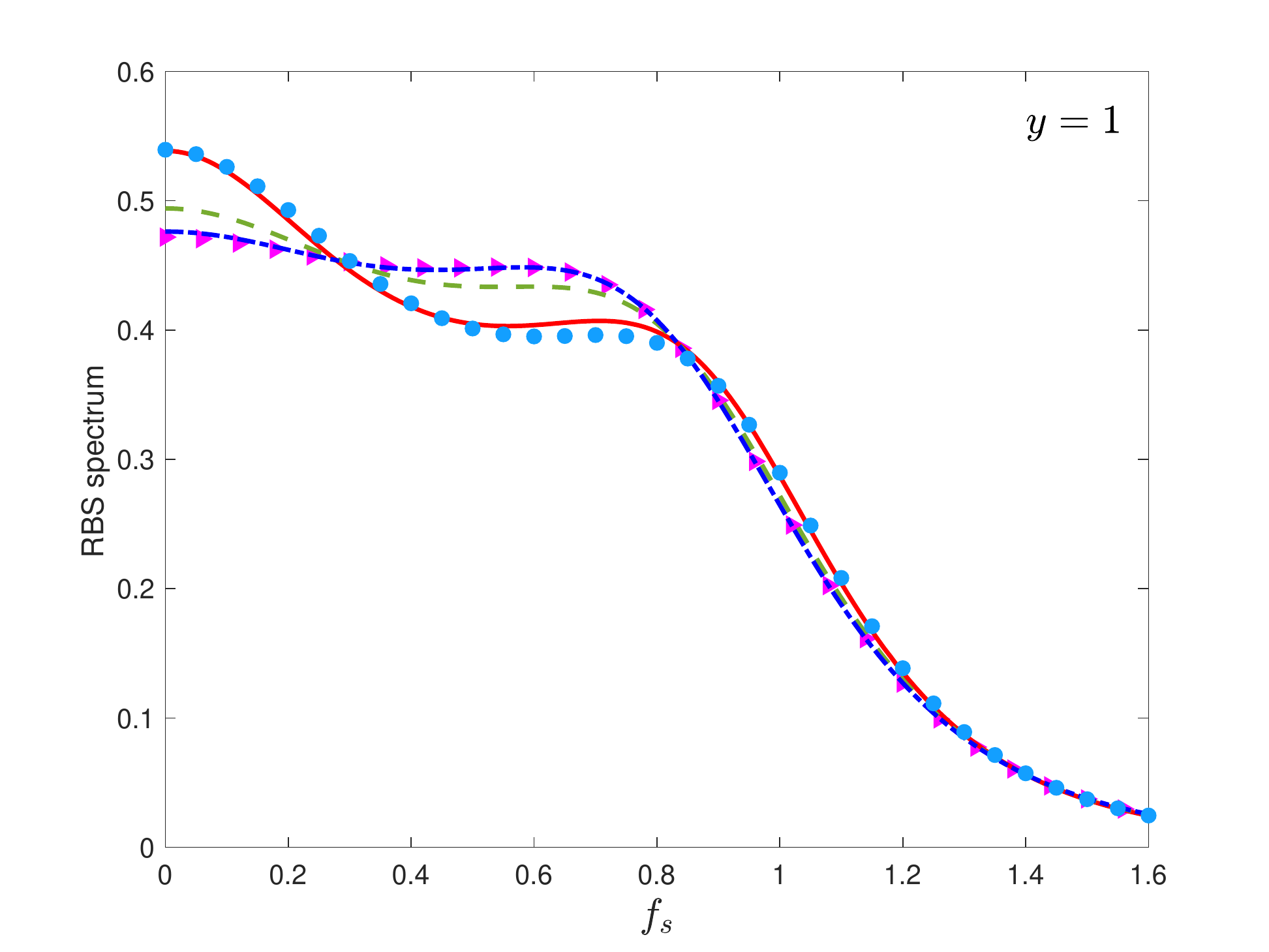}
	\caption{Computed RBS spectra displaying the influence of translational Eucken factor $f_{tr}$ on RBS spectra of molecular gases with $d_r=2$ and $d_v=0$, when the total Eucken factor is $f_u=2$ and the rotational collision number is $Z=2$. In the Wu model~\eqref{SRBS} we change the values of $f_{tr}$ and $f_{int}$ simultaneously according to~\eqref{fint} so that $f_u$ is fixed at 2. In the Tenti-S6 model we have $f_{tr}=2.3077$ according to~\eqref{ftr_Tenti}.  }
	\label{fig:polyd2h1_eucken}
\end{figure}

To this end, we consider the case where the vibrational degrees of freedom are not activated, i.e. $d_v=0$; this occurs in, for example, oxygen at a temperature (e.g. room temperature) far smaller than the characteristic temperature of activation (around 2274~K) of the vibrational energy. We also choose the total Eucken factor $f_u=2$ and the rotational collision number $Z=2$. Thus from \eqref{ftr_Tenti} we know that the translational Eucken factor in the Tenti-S6 model is  $f_{tr}=2.3077$. In the Wu model~\eqref{SRBS}, however, it is possible to vary the values of $f_{tr}$ and $f_{int}$ while keeping $f_u$ fixed to investigate the influence of translational Eucken factor on RBS line shapes. That is, when $f_{tr}$ and $f_u$ are fixed, $f_{int}$ is determined  by
\begin{equation}\label{fint}
f_{int}=\frac{(3+d_r+d_v)f_u-3f_{tr}}{d_r+d_v}.
\end{equation}

The computational results for the kinetic regime are summarized in Figure~\ref{fig:polyd2h1_eucken}. With the same value of $f_{tr}$, the Wu model and Tenti-S6 model produce almost the same line shape. However, in the Wu model the variation of $f_{tr}$ leads to a strong variation of the RBS spectrum, even when the total thermal conductivity is fixed: the RBS spectrum near the Rayleigh peak decreases when $f_{tr}$ increases; the strongest variation occurs around $y\sim1$, and vanishes when $y\gtrsim5$ or $y\lesssim0.2$, i.e. in the hydrodynamic regime and in the near-collisionless regime, respectively. This important fact has not been discovered before, since according to~\eqref{ftr_Tenti}, $f_{tr}$ in the Tenti-S6 model cannot be varied once the bulk viscosity and $f_u$ are fixed. Combining the results in Section \ref{rotational_role} it can be concluded that, in the kinetic regime, if the wrong value of translational Eucken factor is used, gas properties such as the bulk viscosity extracted from the experimental RBS spectrum may be not accurate.

\subsection{Role of effective thermal conductivity}

\begin{figure}
	\centering
	\includegraphics[scale=0.4]{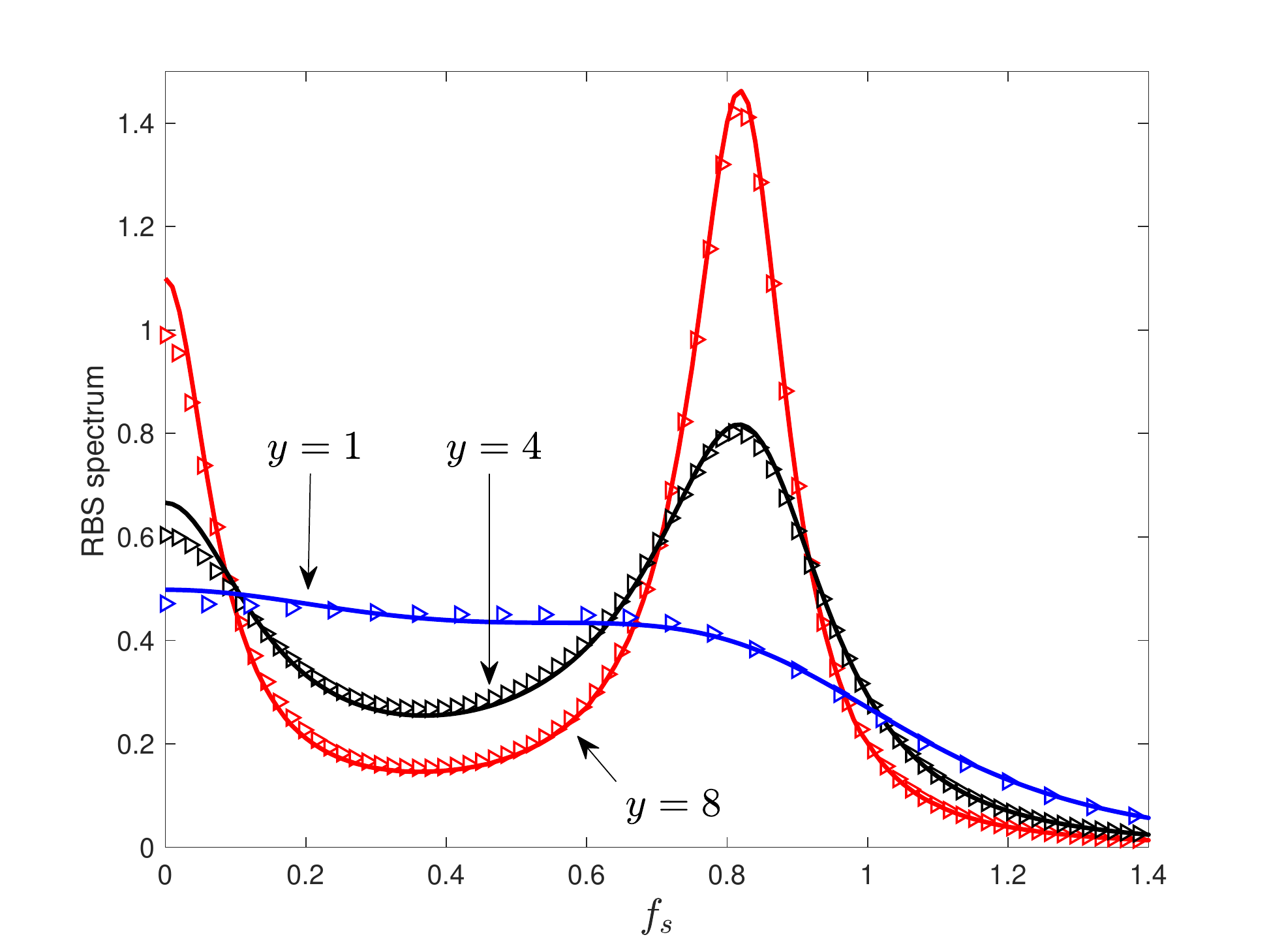}
	\caption{Computed RBS spectra from the Tenti-S6 model with $f_u^e=1.875$ (triangles) and the Wu model~\eqref{SRBS} with $f_u^e=1.667$ and $f_{tr}=2$ for Maxwellian molecules (lines) show the influence of effective Eucken factor $f_u^e$ on RBS spectra of molecular gases with $d_r=3$ and $d_v=15$. The rotational collision number is $Z=3$ in both models, while the calculations were performed for values of the uniformity parameter, $y=1$ (blue), $y=4$ (red), and $y=8$ (black).  }
	\label{fig:effective_eucken}
\end{figure}

The measured value of total Eucken factor in heat conduction experiments contains all contributions from the translational, rotational and vibrational motions of gas molecules. However, in RBS, the vibrational relaxation is ``frozen'', and therefore the contribution of vibrational modes to the total thermal conductivity should be removed when applying the kinetic model to calculate the line shape~\citep{Yang2017CPL,Yang2018Quantitative,WangJ2019CP}. In the Wu model~\eqref{SRBS}, this is not a problem: when $f_u$ and $f_{tr}$ are given, $f_{int}$ can be calculated according to~\eqref{fint}. Then the effective  thermal conductivity  $\kappa_e$ can be calculated from~\eqref{kappa_e}.

In the Tenti-S6 model, however, $\kappa_e$ is determined by setting $f_{tr}=2.5$, while $f_{int}$ is calculated based on the diffusion coefficient~\citep{Yang2017CPL,Yang2018Quantitative,WangJ2019CP}. Since for a molecular gas the real value of $f_{tr}$ is smaller than 2.5, $\kappa_e$ and $f_u^e$ in the Tenti-S6 model are always larger than those in the Wu model. For example, we consider $\operatorname{SF_6}$ with $d_r=3$, $d_v=15$, and $f_u=1.429$. The effective Eucken factor used in the Tenti-S6 model will be $f_u^e=1.875$. However, if we take $f_{tr}=2$, the effective Eucken factor used in the Tenti-S6 model will be $f_u^e=1.667$. This causes the difference in RBS line shapes between the Wu and Tenti-S6 models not only in the kinetic regime (as discussed in Section~\ref{translational_role}), but also in the hydrodynamic regime (i.e. large $y$ values): from  Figure~\ref{fig:effective_eucken} we see that the differences in the RBS spectra at $f_s=0$ between the Wu and Tenti-S6 models are 0.027, 0.065 and 0.114 when the uniformity parameters $y$ are 1, 4 and 8, respectively. If the gas molecule has a smaller value of $f_{tr}$, like for some polar gases such as $\operatorname{CH_3OH}$~\citep{Mason1962}, even larger differences between the Wu and Tenti-S6 models will be found.

\section{Extraction of bulk viscosity and translational Eucken factor}\label{compare_exp}

From the numerical simulations in Section~\ref{GKM_numerical} we see that the RBS line shape is affected by both the bulk viscosity and translational Eucken factor. Here we compare the theoretical RBS line shapes based on the Wu model~\eqref{SRBS} with those from recent experimental measurements, to extract both bulk viscosity and translational Eucken factor.
 
\subsection{$\operatorname{N_2}$}

We first extract the bulk viscosity and translational Eucken factor of $\operatorname{N_2}$ based on the experimental data of~\cite{guziyu}, where the laser wavelength of 366.8~nm is used and the scattering angle is $90^\circ$, i.e. $\theta=\pi/2$. Therefore, the effective wavelength is $L=259.4$~nm. The rotational and vibrational degrees of freedom of $\operatorname{N_2}$ in the experimental condition are $d_r=2$ and $d_v=0$, respectively. The shear viscosity and thermal conductivity are calculated based on the Sutherland formula~\citep{WhiteBook}. The viscosity index $\omega$ is calculated according to~\eqref{viscosity_index} based on Sutherland formula and listed in Table~\ref{table:N2}. The translational and internal Eucken factors in the Wu model~\eqref{SRBS} are chosen to recover the correct value of the thermal conductivity. To determine the rotational collision number $Z$ and the translational Eucken factor $f_{tr}$, we adopt the following procedure: 
\begin{enumerate}
	\item We fix the value of $f_{tr}$, vary the value of $Z$, and  calculate the RBS spectrum based on the Wu model. Then the obtained analytical spectrum is convolved with the instrumental response function to yield the spectrum $S_{WU}(f_s)$. The residual is defined as
	\begin{equation}\label{fitting_error}
	E(Z)=\sum_{\ell=1}^N\frac{\{S_{exp}[f_s(\ell)]-S_{WU}[f_s(\ell)]\}^2}{N},
	\end{equation} 
	where $N$ is the number of discrete frequencies measured in experiments. The minimum residual $E_m(f_{tr})$ is found by fitting $E$ as the quartic polynomial function of $Z$ and finding the minimum of this quartic function. Typically 6 different values of $Z$ are calculated.
	\item Repeat (i) for different values of $f_{tr}$. Typically 6 different values of $f_{tr}$ are calculated. 
	\item Fitting $E_m(f_{tr})$ as the quartic polynomial function of $f_{tr}$ we can determine $f_{tr}$ corresponding to the minimum point of this quartic function. 
	\item With $f_{tr}$ determined in (iii) we do (i) once again to determine $Z$. Hence the bulk viscosity is obtained according to~\eqref{bulk_viscosity}.
\end{enumerate}

\begin{table}
	\centering
	\begin{tabular}{ccccccccc|cc|cc|cc}
& Case & $T$ (K)
&  $10^5\mu_s $
&  $100\kappa $  &  $P$ (bar) &
$y$ & $\omega$ & $\mu_b/\mu_s$ & $f_{tr}$ & $f_{int}$ & $f'_{tr}$ & $f'_{int}$  & $f''_{tr}$ & $f''_{int}$ \\
& (a) &254.7 & $1.57$  & $2.28$ & 2.563 & 1.73 &0.80 &0.48 &2.16 &1.65 &2.17 &1.58 & 2.12 & 1.61  \\
& (b) &275.2 & $1.67$  & $2.44$ & 2.784 & 1.70 &0.78 &0.61 &2.25 &1.53  &2.24 & 1.52 &2.14 &1.59 \\
& (c) &296.7 & $1.77$  & $2.60$ & 3.000 & 1.66 &0.76 &0.69 &2.31 &1.47  &2.27 &1.50 &2.16 &1.58 \\
& (d) & 336.6 & $1.95$ & $2.88$ & 3.400 & 1.61 &0.74 &0.94 &2.43 &1.33 &2.33 &1.45 &2.18 &1.56
\end{tabular}\par
	\caption{Experimental conditions in RBS of $\operatorname{N_2}$~\citep{guziyu} and extracted values of the bulk viscosity and translational/internal Eucken factor based on the Wu model~\eqref{SRBS}. The shear viscosity $\mu_s$ and total thermal conductivity $\kappa$ are shown in SI units. The translational and internal Eucken factors $f'_{tr}$ and $f'_{int}$ are calculated according to~\eqref{Eucken} with ${\rho{D'}}/{\mu_s}=1.32$~\citep{Mason1962}, using the extracted rotational collision number related to the bulk viscosity~\eqref{bulk_viscosity}, while $f''_{tr}$ and $f''_{int}$ are also calculated according to~\eqref{Eucken} but with the rotational collision number calculated from the theory of~\cite{Parker1959}.  }
	\label{table:N2}
\end{table}

\begin{figure}
	\centering
	\includegraphics[scale=0.45,viewport=70 10 850 405,clip=true]{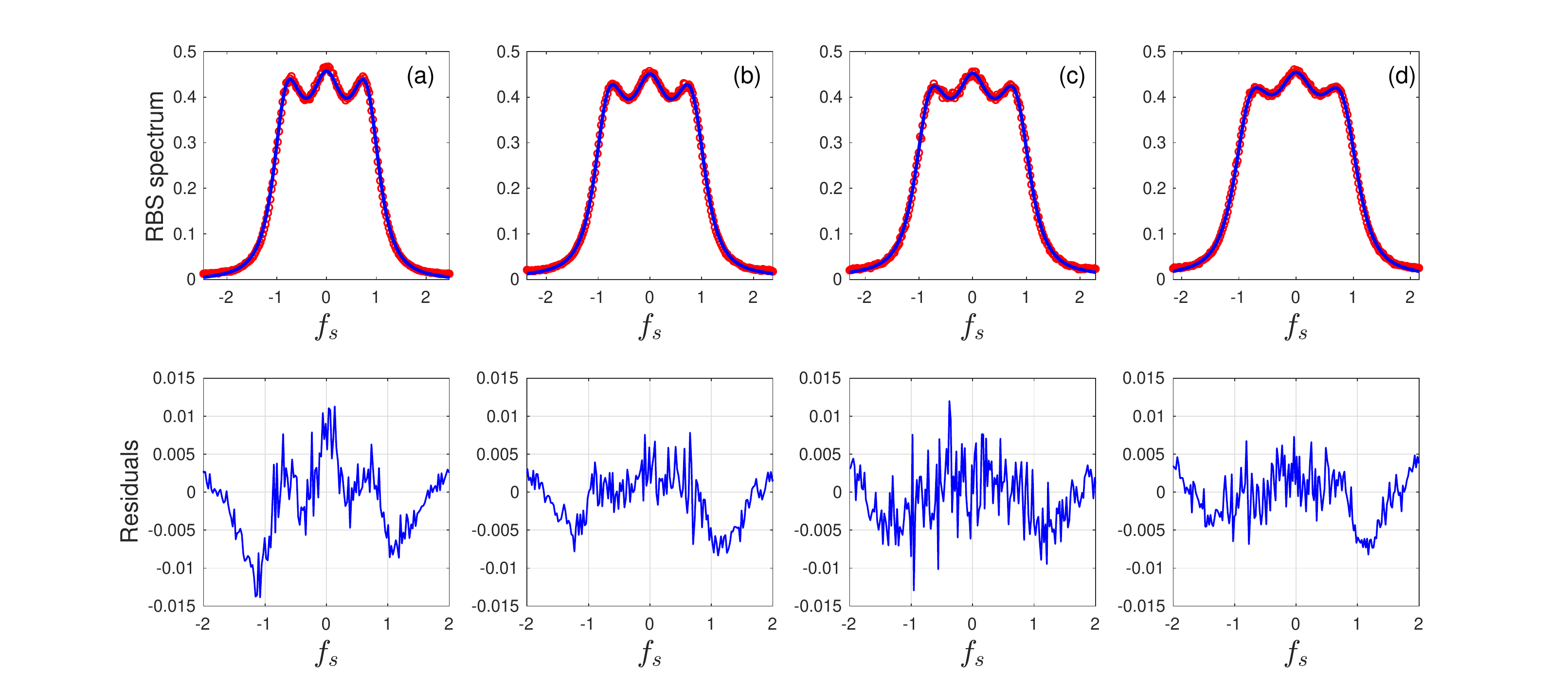}
	\caption{Extraction of the bulk viscosity and translational Eucken factor of $\operatorname{N_2}$ from the experimental RBS spectra $S_{exp}$ (circles) measured by~\cite{guziyu}. Lines in the first row show the RBS spectra $S_{WU}$ obtained from the Wu model with the minimum error~\eqref{fitting_error} over a wide range of $f_{tr}$ and $Z$, while these in the second row show the corresponding residuals between the experimental and theoretical line shapes. Experimental conditions and extracted bulk viscosity and translational/internal Eucken factors are summarized in Table~\ref{table:N2}.   }
	\label{fig:N2}
\end{figure}

Comparisons between the experimental data and the theory of Wu~\eqref{SRBS} are visualized in Figure~\ref{fig:N2}. The residuals between the experimental and theoretical line shapes are generally within 1\%, which are smaller than those by using the Tenti-S6 model, especially for the case (a), see Figure 1 in the paper of~\cite{guziyu}. Note that according to the acoustic experiment~\citep{Lambert1977} the rotational relaxation time of $\operatorname{N_2}$ at standard temperature and pressure is $\tau_r=7.4\times10^{-10}$~s, therefore the bulk viscosity is
\begin{equation}
\begin{aligned}[alignment]
\mu_b=&2p\tau_r\frac{d_r}{(3+d_r)^2}
=2\times10^5\times (7.4\times10^{-10})\times\frac{2}{25}=1.18\times10^{-6}\operatorname{kg\cdot{m^{-1}}\cdot{s^{-1}}},
\end{aligned}
\end{equation}  
which agrees well with the extracted bulk viscosity $\mu_b=1.22\times10^{-5}\operatorname{kg\cdot{m^{-1}}\cdot{s^{-1}}}$ at $T=296.7$~K; for comparison we note that the bulk viscosity extracted by using the Tenti-S6 model is $\mu_b=1.4\times10^{-5}\operatorname{kg\cdot{m^{-1}}\cdot{s^{-1}}}$.

The translational and internal Eucken factors are also extracted from the RBS experiment. We would like to compare these values to the approximate analytical solutions. By making certain assumptions concerning the details of inelastic scattering (i.e. energy exchange between translational and internal modes) and employing a modified Chapman-Enskog expansion, \cite{Mason1962} derived the expressions for translational and internal Eucken factors from the~\cite{WangCS} equation: 
\begin{equation}\label{Eucken}
\begin{aligned}[b]
f_{tr}=\frac{5}{2}\left[1-\frac{10}{3\pi}\left(1-\frac{2\rho{D'}}{5\mu}\right)\left(\frac{d_r}{2Z_{r,coll}}\right)\right], \\
f_{int}=\frac{\rho{D'}}{\mu}\left[ 1+\frac{10}{\pi(d_r+d_v)}\left(1-\frac{2\rho{D'}}{5\mu}\right)\left(\frac{d_r}{2Z_{r,coll}}\right)
\right],
\end{aligned}
\end{equation}
where $Z_{r,coll}=4(d_r+3)Z/3\pi$ is related to the rotational collision time $\tau_r$ as $Z_{r,coll}=4\tau_r/\pi\tau$ with $\tau$ being the mean collision time of gas molecules defined below~\eqref{bulk_viscosity}, and $D'$ is the average diffusion coefficient. Note that it is not easy to determine $Z_{r,coll}$ accurately and $D'$ can be markedly different from the self diffusion coefficient if strong resonant collision occurs~\citep{Mason1962}.  %At room temperature, $Z_{v,coll}$ is several orders larger than $Z_{r,coll}$~\citep{soundexp1959}, therefore the term $h/Z_{v,coll}$ can be neglected unless at extremely high temperatures.

In the paper of~\cite{Mason1962}, ${\rho{D'}}/{\mu}$ takes the value of $1.32$ for nitrogen, for the temperatures listed in Table~\ref{table:N2}. With the extracted rotational collision number $Z$ from the RBS experiment, we can assess the accuracy of analytical expressions for translational and internal Eucken factors. Results are compared with the corresponding extracted values in Table~\ref{table:N2}. The agreements between $f_{tr}$ and $f'_{tr}$, as well as $f_{int}$ and $f'_{int}$ are good, with maximum relative deviation less than 10\%.  We also calculate the analytical translational Eucken factors $f''_{tr}$ and $f''_{int}$ by using the rotational collision number from the theory of~\cite{Parker1959}, and find larger deviations from the experimentally extracted values. Generally speaking, compared to the transport coefficients obtained from the Boltzmann equation for monatomic gas, where the relative deviation is less than about 2\%, analytical transport coefficients of molecular gas~\eqref{Eucken} give less satisfactory agreements with the experimental data, probably due to the introduced approximations in analytical derivations.

\subsection{$\operatorname{CO_2}$}\label{co2_exp}

\begin{figure}
	\centering
	\includegraphics[scale=0.43,viewport=80 330 850 605,clip=true]{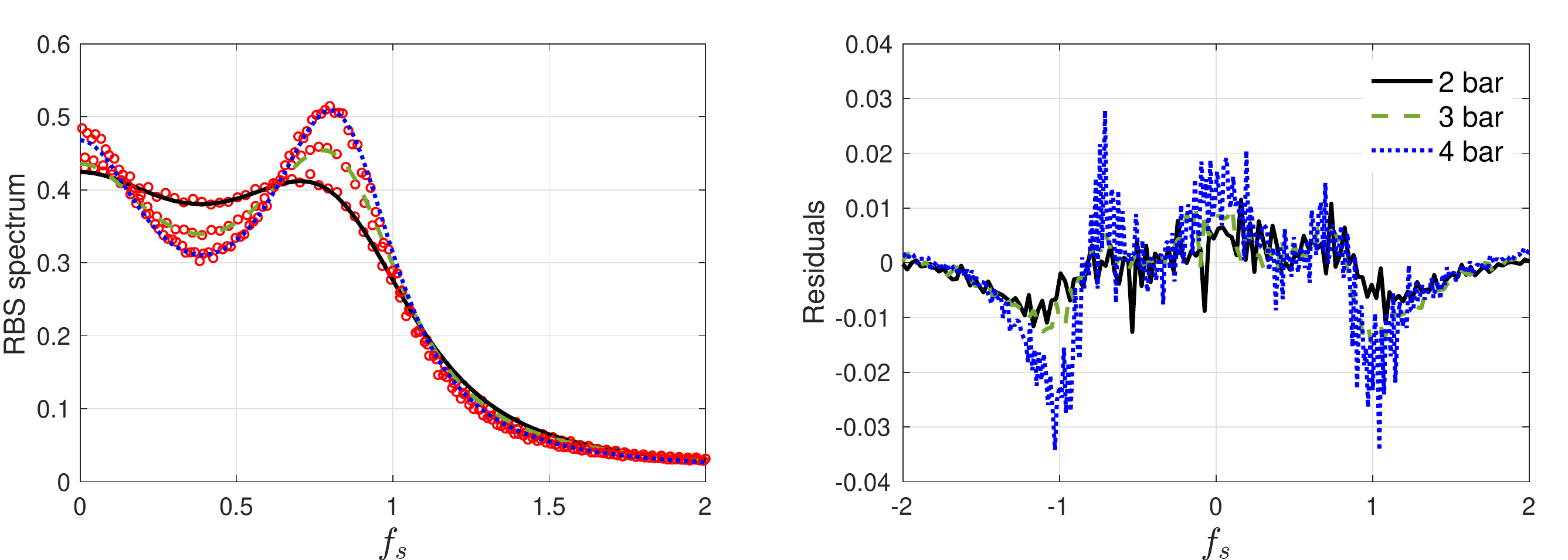}
	\caption{Extraction of the bulk viscosity and translational Eucken factor of $\operatorname{CO_2}$ from the experimental RBS spectra $S_{exp}$ (circles in the left figure) measured by~\cite{Gu2014OL}. The three lines in the left figure show the RBS spectra  obtained from the Wu model. Right figure: residuals between the experimental and theoretical line shapes. The uniformity parameters corresponding to the pressures 2, 3, and 4 bar are $y=1.66$, 2.48, and 3.31, respectively.  }
	\label{fig:CO2}
\end{figure}

We now extract the bulk viscosity and translational Eucken factor of $\operatorname{CO_2}$ based on the experimental data of~\cite{Gu2014OL} at the temperature of 296.5~K. The laser wavelength is 366.8~nm and the scattering angle is $90^\circ$, thus the effective scattering wavelength is $L=259.4$~nm. The shear viscosity is $\mu_s=1.49\times10^{-5}\operatorname{kg\cdot{m^{-1}}\cdot{s^{-1}}}$ and the thermal conductivity is 
$\kappa=1.67\times10^{-2}\operatorname{W\cdot{K^{-1}}\cdot{m^{-1}}}$. The viscosity index used in the Boltzmann equation is $\omega=0.933$~\citep{CE}, the rotational degrees of freedom is  $d_r=2$, and the vibrational  degrees of freedom is $d_v=1.87$ using the data from the National Institute of Standards and Technology. Comparisons between the experimental and theoretical RBS line shapes are shown in Figure~\ref{fig:CO2}, where the extracted rotational collision numbers are $Z=1.72$, 1.56, 1.64, the translational Eucken factors are $f_{tr}=2.30$, 2.23, and 2.18, and the internal Eucken factors are $f_{int}=1.28$, 1.34, and 1.37, at pressures 2, 3, and 4 bar, respectively. We do not consider the case of 1 bar pressure because the uniformity parameter is $y=0.83$ so that the line shape is not sensitive to $f_{tr}$ and $Z$. Taking the average value we find that the translational and internal Eucken factor are $f_{tr}=2.24$ and $f_{int}=1.33$, respectively. The average value for the rotational collision number is $Z=1.60$ and hence the bulk viscosity is $\mu_b=6.6\times10^{-6}\operatorname{kg\cdot{m^{-1}}\cdot{s^{-1}}}$. Note that the rotational relaxation time of $\operatorname{CO_2}$ at standard temperature and pressure is $\tau_r=3.8\times10^{-10}$~s~\citep{Lambert1977}, therefore the bulk viscosity is
\begin{equation}
\begin{aligned}[alignment]
\mu_b=2\times10^5\times (3.8\times10^{-10})\times\frac{2}{25}=6.1\times10^{-6}\operatorname{kg\cdot{m^{-1}}\cdot{s^{-1}}},
\end{aligned}
\end{equation}
which agrees well with the bulk viscosity extracted from the RBS experiment~\citep{Gu2014OL} using the Wu model~\eqref{SRBS}. Note that this bulk viscosity only takes into account the contribution from rotational relaxation of gas molecules; it is smaller than that obtained from the acoustic experiment by about four orders of magnitude~\citep{Pan2005,Lambert1977}.

\subsection{$\operatorname{SF_6}$}

\begin{figure}
	\centering
	\includegraphics[scale=0.43,viewport=80 60 850 605,clip=true]{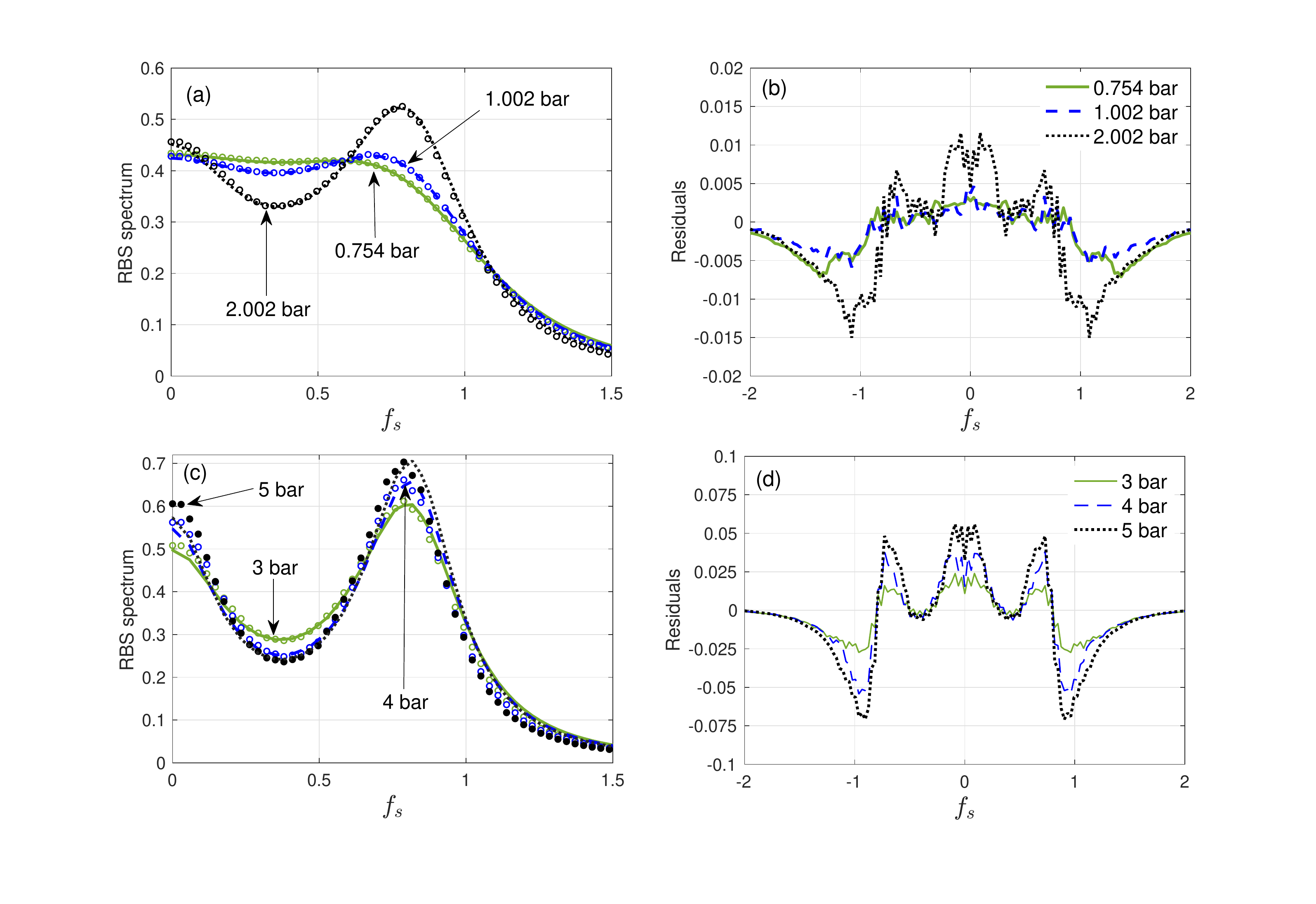}
	\caption{Same as Figure~\ref{fig:CO2}, except the experimental data of $\operatorname{SF_6}$ measured by~\cite{Yang2017CPL} is used here. The uniformity parameters corresponding to the pressures 0.754, 1.002, 2.002, 3, 4, and 5 bar are $y=1.22$, 1.64, 3.26, 4.88, 6.49, and 8.14, respectively.}
	\label{fig:SF6}
\end{figure}

The experimental data of~\cite{Yang2017CPL} for SF$_6$ molecules recorded at pressures 0.754, 1.002, 2.002, 3, 4, and 5 bar are compared to the Wu model, where the laser wavelength is 403~nm and the scattering angle is $89.6^\circ$, so the effective wavelength is $L=286$~nm. With the viscosity data given by \cite{SF6Data} we find that the shear viscosity is $\mu_s=1.52\times10^{-5} \operatorname{kg\cdot{}m^{-1}\cdot{}s^{-1}}$ and the viscosity index is $\omega=0.885$ at temperature 298 K, while the heat conductivity is $\kappa=1.30\times10^{-2} \operatorname{W\cdot{}m^{-1}\cdot{}K^{-1}}$. The rotational degrees of freedom is $d_r=3$, while from the heat capacity data of~\cite{Guder2009} we have $d_v=15.32$. Comparisons between the experimental and theoretical RBS line shapes are shown in Figure~\ref{fig:SF6}, where the extracted rotational collision numbers are $Z=1.85$, 1.81, 1.90, 2.25, 2.64, and 3.31, while the translational Eucken factors are $f_{tr}=2.29$, 2.23, 2.10, 2.07, 1.80, and 2.07, at pressures 0.754, 1.002, 2.002, 3, 4, and 5 bar, respectively. It is noted from the second row in Figure~\ref{fig:SF6} that the residuals between experimental and theoretical data increase significantly with pressure; it is postulated by~\cite{Yang2017CPL} that this might be due to the dense gas effect. However, the simulation results based on the Enskog-Vlasov equation, which takes into account the dense gas effect, still produce line shapes that have large deviations from the RBS experimental data, see Figure 8 in the paper by~\cite{Bruno2019CPL}. We leave this to future investigation and in the present paper we consider the extracted gas thermodynamic properties for pressure below 3 bar. Taking the average value at pressures of 0.754, 1.002, and 2.002 bars we have $f_{tr}=2.21$ and $f_{int}=1.28$. The average value for $Z$ is $1.85$, hence the ratio of bulk to shear viscosities is 0.62. Note that the rotational relaxation time of $\operatorname{SF_6}$ at standard temperature and pressure is $\tau_r=6\times10^{-10}$~s~\citep{Haebel1968}, therefore the viscosity ratio is
\begin{equation}
\begin{aligned}[alignment]
\frac{\mu_b}{\mu_s}=\frac{2\times10^5\times (6\times10^{-10})\times\frac{3}{36}}{1.52\times10^{-5}}=0.66,
\end{aligned}
\end{equation}
which agrees with the extracted value from the RBS experiment~\citep{Yang2017CPL}. Note that again this bulk viscosity only takes into account the contribution from the rotational relaxation of gas molecules.

\section{Conclusions}\label{conclusion}

In summary, we have analysed how an RBS spectrum in the kinetic regime (when the uniformity parameter falls in the region $1.2\precapprox{}y\precapprox3.3$) is sensitive to both the bulk viscosity and the translational Eucken factor and how the effect of these thermodynamic gas properties can be disentangled and determined separately from experiment. The Wu model is used to extract the bulk viscosity (due to the rotational relaxation of gas molecules only) and translational Eucken factor from recent RBS experiments of $\operatorname{N_2}$, $\operatorname{CO_2}$ and $\operatorname{SF_6}$, which has two advantages over the prevailing Tenti-S6 model: (i) the bulk viscosity and translational Eucken factor can be determined independently and (ii) the Boltzmann equation can be recovered in the limit of infinite rotational relaxation time. The extracted bulk viscosity agrees well with the measurements from acoustic experiments when only the rotational relaxation is considered. For the extraction of the translational Eucken factor, our method using the RBS line shapes is unique. This is because historically the translational Eucken factor is measured in thermal creep flows in the kinetic regime, but this measurement approach is essentially hampered by gas-surface boundary effects. This problem is entirely absent in RBS experiments, since only a local volume inside the gas cell is probed by the laser light so the gas-surface interaction is absent.

Although the extracted value of translational Eucken factor has no application in hydrodynamic flows (where only the total thermal conductivity is important), it will affect the heat transfer in rarefied gas flows, as the rarefaction effects corresponding to the translational and internal motions are different due to the difference between the mean collision time of gas molecules and the rotational/vibrational relaxation times. Thus, when analysing the results of RBS spectral measurements in the context of presented theoretical approach, the RBS experiments provide critical information on translational and internal Eucken factors. This allows for more reliable simulations of rarefied gas dynamics, especially for polar gases where the translational Eucken factor could be much smaller than the value 2.5 of monatomic gases.
  
\newpage
 
\bibliographystyle{jfm}
%\bibliography{bibnew}

%\begin{thebibliography}{99}

\end{document}